\documentclass[format=acmsmall]{acmart}

\usepackage{booktabs} 
\usepackage{comment}
\usepackage{subfigure}
\usepackage{pifont}
\usepackage{graphicx}
\usepackage{color}
\usepackage{ifthen}
\usepackage{multirow}
\usepackage{algorithm}
\usepackage{algorithmic}
\usepackage{balance}
\usepackage{url}
\usepackage{amsmath}

\newcommand{\dif}{\mathrm{d}} 

\newcommand{\sect}[1]{\textsection#1}

\DeclareMathOperator*{\argmin}{arg\,min}

\usepackage{ifluatex}
\ifluatex
\usepackage{hyperref}
\usepackage{spelling}
\spellingreadgood{/misc/tools/dictionaries/whitelist}
\spellingreadgood{/misc/tools/dictionaries/whitelist2}
\fi

\copyrightyear{2009} 

\setcopyright{acmcopyright}
\acmJournal{TOSN}
\acmYear{2019} 
\acmVolume{1} 
\acmNumber{1} 
\acmArticle{1} 
\acmMonth{1} 
\acmPrice{15.00}

\acmDOI{0}

\received{July 2018}
\received[revised]{March 2009}
\received[accepted]{June 2009}

\begin{document}

\title{Attack-Aware Synchronization-Free Data Timestamping in LoRaWAN}

\author{Chaojie Gu}
\email{gucj@ntu.edu.sg}
\author{Linshan Jiang}
\email{LINSHAN001@e.ntu.edu.sg}
\author{Rui Tan}
\email{tanrui@ntu.edu.sg}
\author{Mo Li}
\email{limo@ntu.edu.sg}
\affiliation{%
  \institution{Nanyang Technological University}
  \department{School of Computer Science and Engineering}
  \streetaddress{N4-B02A-01, 50 Nanyang Avenue}
  \country{Singapore} 
  \postcode{639798}
}

\author{Jun Huang}
\affiliation{%
  \institution{Massachusetts Institute of Technology}
  \department{Sloan School of Management}
  \streetaddress{100 Main St, Cambridge}
  \country{United States} 
  \postcode{MA 02142}
}
\email{junhuang@mit.edu}

\begin{abstract}
  Low-power wide-area network technologies such as LoRaWAN are promising for collecting low-rate monitoring data from geographically distributed sensors, in which timestamping the sensor data is a critical system function. This paper considers a synchronization-free approach to timestamping LoRaWAN uplink data based on signal arrival time at the gateway, which well matches LoRaWAN's one-hop star topology and releases bandwidth from transmitting timestamps and synchronizing end devices' clocks at all times. However, we show that this approach is susceptible to a {\em frame delay attack} consisting of malicious frame collision and delayed replay. Real experiments show that the attack can affect the end devices in large areas up to about $50,000\,\text{m}^2$. In a broader sense, the attack threatens any system functions requiring timely deliveries of LoRaWAN frames. To address this threat, we propose a $\mathsf{LoRaTS}$ gateway design that integrates a commodity LoRaWAN gateway and a low-power software-defined radio receiver to track the inherent frequency biases of the end devices. Based on an analytic model of LoRa's chirp spread spectrum modulation, we develop signal processing algorithms to estimate the frequency biases with high accuracy beyond that achieved by LoRa's default demodulation. The accurate frequency bias tracking capability enables the detection of the attack that introduces additional frequency biases. We also investigate and implement a more crafty attack that uses advanced radio apparatuses to eliminate the frequency biases. To address this crafty attack, we propose a pseudorandom interval hopping scheme to enhance our frequency bias tracking approach. Extensive experiments show the effectiveness of our approach in deployments with real affecting factors such as temperature variations.
\end{abstract}

%
%

\renewcommand{\shortauthors}{C. Gu et al.}


\thanks{A preliminary version of this work appears in The 40th IEEE International Conference on Distributed Computing Systems (ICDCS 2020). This research was supported in part by two MOE AcRF Tier 1 grants (2019-T1-001-044 and 2018-T1-002-081).}

\maketitle

\section{Introduction}
\label{sec:intro}
Low-power wide-area networks (LPWANs) enable direct wireless interconnections among end devices and gateways in geographic areas of square kilometers. It increases network connectivity as a defining characteristic of the Internet of Things (IoT). Among various LPWAN technologies (including NB-IoT and Sigfox), LoRaWAN, which is an open data link layer specification based on the LoRa modulation scheme \cite{lorawan-spec}, offers the advantages of using license-free ISM bands, low costs for end devices, and independence from managed cellular infrastructures.

LoRaWAN is promising for the applications of collecting low-rate monitoring data from geographically distributed sensors, such as utility meters, environment sensors, roadway detectors, industrial measurement devices, etc. 
All these applications require data timestamping as a basic system service, though they may require different timestamp accuracies. 
For instance, data center environment condition monitoring generally requires sub-second accuracy for sensor data timestamps to capture the thermodynamics \cite{chen2012high}. Sub-second-accurate timestamps for the traffic data generated by roadway detectors can be used to reconstruct real-time traffic maps \cite{oh2002real}. In a range of industrial monitoring applications such as oil pipeline monitoring, milliseconds accuracy may be required \cite{pipeline}. In volcano monitoring, the onset times of seismic events detected by geographically distributed sensors require sub-10 milliseconds accuracy to be meaningful to volcanic earthquake hypocenter estimation \cite{liu2013volcanic}.

There are two basic approaches, namely, {\em sync-based} and {\em sync-free}, to data timestamping in wireless sensor networks (WSNs). In the sync-based approach, the sensor nodes keep their clocks synchronized and use the clock value to timestamp the data once generated. Differently, the sync-free approach uses the gateway with wall time to timestamp the data upon the arrival of the corresponding network frame. Based on various existing distributed clock synchronization protocols, multi-hop WSNs mostly adopt the sync-based approach. The sync-free approach is ill-suited for multi-hop WSNs, because the data delivery on each hop may have uncertain delays due to various factors such as channel contention among nodes.

In contrast, LoRaWANs prefer the sync-free approach for uplink data timestamping. Reasons are two-fold. First, different from multi-hop WSNs, LoRaWANs adopt a one-hop gateway-centered star topology that is free of the issue of hop-wise uncertain delays. Specifically, as the radio signal propagation time from an end device to the gateway is generally in microseconds, the LoRaWAN frame arrival time can well represent the time when the frame leaves the end device. As a result, timestamping the uplink data at the gateway can meet the milliseconds or sub-second timestamping accuracy requirements of many applications. Second, if the sync-based approach is adopted otherwise, the task of keeping the end devices' clocks synchronized at all times and the inclusion of timestamps in the LoRaWAN data frames will introduce communication overhead to the narrowband LoRaWANs (a detailed analysis can be found in \sect\ref{subsec:sync-vs-free}). Therefore, performance-wise, the sync-free approach well matches LoRaWANs' star topology and addresses its bandwidth scarcity.

However, LoRaWAN's long-range communication capability also renders itself susceptible to wireless attacks that can be launched from remote and hidden sites. The attacks may affect many end devices in large geographic areas. In particular, the conventional security measures that have been included in the LoRaWAN specifications (e.g., frame confidentiality and integrity) may be inadequate to protect the network from wireless attacks on the physical layer. Therefore, it is of importance to study the potential wireless attacks against the sync-free data timestamping, since inaccurate and even incorrect timestamps significantly undermine the value of the data. For example, when applying LoRa for transmitting detected earthquake events, tiny timestamping errors will lead to earthquake hypocenter estimation errors \cite{liu2013volcanic}. In this paper, we consider a basic threat of {\em frame delay attack} that directly invalidates the assumption of near-zero signal propagation time. Specifically, by setting up a {\em collider} device close to the LoRaWAN gateway and an {\em eavesdropper} device at a remote location, a combination of malicious frame collision and delayed replay may introduce arbitrary delays to the deliveries of uplink frames. Although wireless jamming and replay have been studied extensively, how easily they can be launched in a coordinated manner to introduce frame delay and how much impact (e.g., in terms of the affected area) the attack can generate are still open questions in the context of LoRaWANs.

This paper answers these questions via real experiments. Our measurements show that LoRa demodulators have lengthy vulnerable time windows, in which the gateway cannot decode either the victim frame or the collision frame, and raises no alerts. Thus, it is easy to launch stealthy attacks by exploiting the vulnerable time windows. In particular, as the attack does not breach the integrity of the frame content and sequence, the attack cannot be solved by cryptographic protection and frame counting. Our experiments in a campus LoRaWAN show that, a fixed setup of a collider and an eavesdropper can subvert the sync-free data timestamping service for end devices in a large geographic area of about $50,000\,\text{m}^2$. In a broader sense, this attack threatens any system functions that require timely deliveries of uplink frames in LoRaWAN. Note that this attack is valid but marginally important in short-range wireless networks (e.g., Zigbee and Wi-Fi) because of the limited area affected by the attack and the difficulty in controlling the attack radios' timing. Differently, it is important to LoRaWANs because it can affect large geographic areas, and the timing of the attack radios can be easily controlled due to LoRaWAN's long symbol times.

Therefore, an upgraded sync-free timestamping approach that integrates countermeasures against the attack and meanwhile preserves the bandwidth efficiency is desirable. Moreover, it should only require changes to the gateway. In this paper, we aim to develop an awareness of the attack by monitoring the end devices' radio frequency biases (FBs). Due to the manufacturing imperfections of the radio chips' internal oscillators, each radio chip generally has an FB that is the difference between the frequency of the carrier signal emitted by the chip and the nominal value. A change of FB detected by the gateway suggests the received frame may be a replayed one since the adversary's replay device superimposes its own FB onto the replayed signal. To access the physical layer, we integrate a low-cost (US\$25) software-defined radio (SDR) receiver \cite{rtl-sdr} with a commodity LoRaWAN gateway to form our LoRa TimeStamping ($\mathsf{LoRaTS}$) gateway. We develop time-domain signal processing algorithms for $\mathsf{LoRaTS}$ to estimate the FB. Experiments show that (i) with a received signal-to-noise ratio (SNR) of down to $-18\,\text{dB}$, $\mathsf{LoRaTS}$ achieves an accuracy of $120\,\text{Hz}$ in estimating FB, which is just 0.14 parts-per-million (ppm) of the channel's central frequency of 869.75 MHz; (ii) the frame replay by an SDR transceiver introduces an additional FB of at least 0.24 ppm. Thus, $\mathsf{LoRaTS}$ can track FB to detect the replay step of the frame delay attack. 
Note that the detection does not require uniqueness or distinctiveness of the FBs across different LoRa transceivers because it is based on changes of FB.

In summary, $\mathsf{LoRaTS}$ supports the bandwidth-efficient sync-free timestamping and requires no modifications on the LoRaWAN end devices. It is a low-cost countermeasure that increases the cost and technical barrier for launching effective frame delay attacks since the attackers need to eliminate the tiny FBs of their radio apparatuses. $\mathsf{LoRaTS}$ strikes a satisfactory trade-off between network efficiency and the security level required by typical LoRaWAN applications. If the attackers employ expensive radio apparatuses to eliminate the tiny FBs, we further develop an approach based on the pseudorandom number generator to counteract the zero-FB attack. With the zero-FB attack countermeasure deployed, although the end device needs to follow a transmission schedule determined by the pseudorandom number generator, the security level of the system is further improved.

This paper makes the following contributions:
\begin{itemize}
    \item We implement the frame delay attack against LoRaWANs. Simulations and experiments show the large sizes of the geographic areas vulnerable to the attack.
    \item Based on an analytic model of LoRa's chirp spread spectrum (CSS) modulation, we design a time-domain signal processing pipeline to accurately estimate end devices' FBs.
    \item Extensive experiments in indoor and outdoor environments show that $\mathsf{LoRaTS}$ can detect the frame delay attacks that introduce additional FBs.
    \item We implement the zero-FB attack and propose a Pseudorandom Interval Hopping scheme to counteract the zero-FB attack. 
\end{itemize}

The rest of this paper is organized as follows. \sect\ref{sec:related} reviews related work; \sect\ref{sec:timestamping} describes sync-free data timestamping; \sect\ref{sec:security} studies the attack;
\sect\ref{sec:lorasync} presents $\mathsf{LoRaTS}$ and uplink frame arrival time detection approach;
\sect\ref{sec:fingerprint} studies LoRa's FB and uses it to detect attack;
\sect\ref{sec:zero-fb} studies zero-FB attack and countermeasure;
\sect\ref{sec:eval} presents experiment results;
\sect\ref{sec:conclude} concludes this paper.

\section{Related Work}
\label{sec:related}

Improving LoRaWAN's communication performance has received increasing research.
The Choir system \cite{eletreby2017empowering} exploits the diverse FBs of the LoRaWAN end devices to disentangle colliding frames from different end devices. Choir uses the dechirping and Fourier transform processing pipeline to analyze FB, which does not provide sufficient resolution for detecting the tiny extra FB introduced by attack (see details in \sect\ref{subsubsec:extraction}). In this paper, based on an analytic model of LoRa's CSS modulation, we develop a new time-domain signal processing algorithm based on a least squares formulation to achieve the required resolution.
The Charm system \cite{charm2018} exploits \textit{coherent combining} to decode a frame from the weak signals received by multiple geographically distributed LoRaWAN gateways. It allows the LoRaWAN end device to use a lower transmitting power. Several recent studies \cite{peng2018plora,hessar19} have devised various backscatter designs for LoRa to reduce the power consumption of end devices.
All the studies mentioned above focus on understanding and improving the data communication performance of LoRaWAN \cite{eletreby2017empowering,charm2018}, or reducing power consumption via backscattering \cite{peng2018plora,hessar19}. None of them specifically addresses efficient data timestamping, which is a basic system function of many LoRaWAN-based systems.

LongShoT \cite{ramirez2019longshot} is an approach to synchronize the LoRaWAN end devices with the gateway. Through low-level offline time profiling for a LoRaWAN radio chip (e.g., to measure the time delays between hardware interrupts and the chip's power consumption rise), LongShoT achieves sub-50 microseconds accuracy, which is echoed by our results on the accuracy of estimating signal arrival time using a different approach. LongShoT is designed for the LoRaWAN systems requiring tight clock synchronization. Differently, we address data timestamping and focus on the less stringent but more commonly seen milliseconds or sub-second accuracy requirements. Our sync-free approach releases the bandwidth from frequent clock synchronization operations.

Security of LoRaWAN is receiving research attention. In \cite{7985777}, Aras et al. discuss several possible attacks against LoRaWAN, including key compromise and jamming. The key compromise requires prior physical attack of memory extraction.
In \cite{selective-jamming}, a selective jamming attack against certain receivers and/or certain application frames is studied. 
Different from the studies \cite{7985777, selective-jamming} that do not consider the stealthiness of jamming, we consider stealthy frame collision.
From our results in \sect\ref{subsubsec:experiments}, the selective jamming in \cite{selective-jamming} cannot be stealthy because it cannot start jamming until the frame header is decoded and the corruption of payload must lead to integrity check failures. In \cite{robyns2017physical}, Robyns et al. apply supervised machine learning for end device classification based on the received LoRa signal.
From our measurements, the dissimilarity between the original and the replayed signals is much lower than that among the original signals from different end devices. Thus, the supervised machine learning is not promising for attack detection. 

Device identification based on radiometric features has been studied for short-range wireless technologies.
A radiometric feature is the difference between the nominal and the measured values of a certain modulation parameter. 
The work \cite{brik2008wireless} studied the radiometric features of IEEE 802.11 radios, including symbol-level features regarding signal magnitude and phase, as well as the frame-level feature regarding carrier frequency. 
In LoRaWAN, the received signal strength is often rather low due to long-distance propagation or barrier penetration. 
As such, the signal magnitude radiometric feature cannot be used as a radiometric feature. 
As the phase of LoRa signal is arbitrary, it cannot be employed as a radiometric feature too. 
In this paper, we show that the bias of the LoRa signal's carrier frequency from the nominal value is an effective radiometric feature. 
This feature can be used to counteract the frame delay attack. Based on LoRa's CSS modulation, we develop a lightweight algorithm that can estimate this feature from the received LoRa signal. It requires a low-cost SDR receiver, unlike the expensive vector signal analyzer \cite{vectoranalyzer} used in the work \cite{brik2008wireless}.

\section{Data Timestamping in LoRaWAN}
\label{sec:timestamping}

\subsection{LoRaWAN Primer}
LoRa is a physical layer technique that adopts CSS modulation. LoRaWAN is an open data link specification based on LoRa. A LoRaWAN is a star network consisting of a number of {\em end devices} and a {\em gateway} that is often connected to the Internet. Gateways are often equipped with GPS receivers for time keeping. The transmission direction from the end device to the gateway is called {\em uplink} and the opposite is called {\em downlink}. LoRaWAN defines three classes for end devices, i.e., Class A, B and C. In Class A, each communication session must be initiated by an uplink transmission. There are two subsequent downlink windows.
Class A end devices can sleep to save energy when there are no pending data to transmit. Class A adopts the ALOHA media access control protocol.
Class B extends Class A with additional scheduled downlink windows. 
However, such scheduled downlink windows require the end devices to have synchronized clocks, incurring considerable overhead as we will analyze shortly. 
Class C requires the end devices to listen to the channel all the time. Clearly, Class C is not for low-power end devices.
In this paper, we focus on Class A, because it is supported by all commodity platforms and energy-efficient. To the best of our knowledge, no commodity platforms have out-of-the-box support for Class B that requires clock synchronization.

\subsection{Advantages of Sync-Free Timestamping}
\label{subsec:sync-vs-free}

Data timestamping, i.e., to record the {\em time of interest} in terms of the wall clock, is a basic system function required by the data collection applications for monitoring. 
For a sensor measurement, the time of interest is the time instant when the measurement is taken by the end device. Multi-hop WSNs largely adopt the {\em sync-based} approach. Specifically, the clocks of the WSN nodes are synchronized to the global time using some clock synchronization protocol. Then, each WSN node can timestamp the data using its local clock. WSNs have to adopt this approach due primarily to that the multi-hop data deliveries from the WSN nodes to the gateway in general suffer uncertain delays. Thus, although the clock synchronization introduces additional complexity to the system implementation, it has become a standard component for systems requiring data timestamping. However, the clock synchronization introduces considerable communication overhead to the bandwidth-limited LoRaWANs.

We present an example to illustrate the overhead to maintain sub-10 milliseconds (ms) clock accuracy in LoRaWANs.
Typical crystal oscillators in microcontrollers have drift rates of $30$ to $50\,\text{ppm}$ \cite{wizsync}. Without loss of generality, we adopt $40\,\text{ppm}$ for this example. With this drift rate, an end device needs 14 synchronization sessions per hour to maintain sub-$10\,\text{ms}$ clock accuracy. These 14 sessions represent a significant communication overhead for an end device. For instance, in Europe, a LoRaWAN end device adopting a spreading factor of 12 can only send 24 30-byte frames per hour to conform to the 1\% duty cycle requirement \cite{tech-LPWAN}. Although the synchronization information may be piggybacked to the data frames, a low-rate monitoring application may have to send the frames more frequently just to keep time. In addition, the data frames need to include data timestamps, each of which needs at least a few bytes. This is also an overhead given the bandwidth scarcity. 

To efficiently utilize LoRaWAN's scarce bandwidth and exploit its star topology, the sync-free timestamping approach can be adopted. In this approach, an end device transmits a sensor reading once generated. Upon receiving the frame, the gateway uses the frame arrival time as the data timestamp. The signal propagation time from the end device to the gateway, which is often microseconds, can be ignored for millisecond-accurate timestamping. Compared with the sync-based approach, this sync-free approach avoids the communication overhead caused by the frequent clock synchronization operations and the transmissions of timestamps. Thus, the sync-free approach is simple and provides bandwidth-saving benefit throughout the lifetime of the LoRaWANs.

\section{Security of Sync-Free Timestamping}
\label{sec:security}

The long-range communication capability of LoRaWAN enables the less complex and bandwidth-efficient sync-free timestamping. However, it may also be subject to wireless attacks that can affect large geographic areas.
Having understood the benefit of sync-free timestamping, we also need to understand its security risk and the related countermeasure for achieving a more comprehensive assessment on the efficiency-security tradeoff.
A major and direct threat against the sync-free approach is the {\em frame delay attack} that manipulates the frame delivery time to invalidate the assumption of near-zero signal propagation delay. We define the attack as follows.

\noindent {\bf Frame delay attack:} The end device and gateway are not corrupted by the adversary. However, the adversary may delay the deliveries of the uplink frames. The malicious delay for any uplink frame is finite. Moreover, the frame cannot be tampered with because of cryptographic protection.

The attack results in wrong timestamps under the sync-free approach. This section studies the attack implementation (\sect\ref{subsec:attack-implementation}), investigates the timing of malicious frame collision (\sect\ref{subsubsec:experiments}), and studies the size of the vulnerable area in which the end devices are affected by the attack (\sect\ref{subsec:attack-surface}).

\subsection{Attack Implementation}
\label{subsec:attack-implementation}

\subsubsection{Implementation steps}
\label{subsubsec:implementation-principle}

\begin{figure}
	\centering
	\includegraphics[width=.8\textwidth]{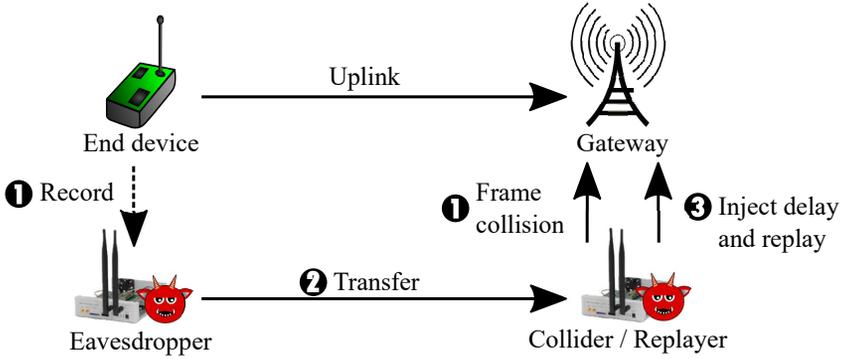}
	\caption{Steps for implementing frame delay attack.}
	\label{fig:lora_p2p_attack}
\end{figure}

Fig.~\ref{fig:lora_p2p_attack} illustrates the attack implementation. The adversary sets up two malicious devices called {\em eavesdropper} and {\em collider} that are close to the end device and the gateway, respectively. The attack consists of three steps. \ding{182} At the beginning, both the eavesdropper and the collider listen to the LoRa communication channel between the end device and the gateway. Once the collider detects an uplink frame transmission, it transmits a collision frame. In \sect\ref{subsubsec:experiments}, we will investigate experimentally a stealthy collision method such that the victim gateway does not raise any warning message to the application layer. Meanwhile, once the eavesdropper detects an uplink frame transmission, it records the radio waveform of the frame. Note that the collider may choose a proper transmitting power of the collision frame such that the collision can affect the victim gateway, while not corrupting the radio waveform recorded by the eavesdropper.
\ding{183} The eavesdropper sends the recorded radio waveform data to the collider via a separate communication link that provides enough bandwidth. \ding{184} After a time duration of $\tau$ seconds from the onset time of the victim frame transmission, the collider replays the recorded radio waveform. Thus, in this paper, the {\em collider} and the {\em replayer} refer to the same attack device. The above collision-and-replay process does not need to decipher the payload of the recorded frame; it simply re-transmits the recorded radio waveform. As the gateway cannot receive the original frame and the integrity of the replayed frame is preserved, the gateway accepts the replayed frame even if it checks the frame integrity and frame counter. The attack introduces a delay of $\tau$ seconds to the delivery of the frame.

We discuss several issues in the attack implementation. 
First, using a normal LoRaWAN frame to create malicious collision is more stealthy than brute-force jamming, since it may be difficult to differentiate malicious and normal collisions. 
Brute-force jamming can be easily detected and located. Second, as the adversary delays the uplink frame, how does the adversary know in time the direction of the current transmission? In LoRaWAN, the uplink preamble uses up chirps, whereas the downlink preamble uses down chirps. Thus, the adversary can quickly detect the direction of the current transmission within a chirp time. From our results in \sect\ref{subsubsec:experiments}, a time duration of one chirp for sensing the direction of the transmission does not impede the timeliness of the collision attack. 
Third, to increase the stealthiness of the replay attack, the replayer can well control the transmitting power of the replay such that only the victim gateway can receive the replayed frame. Fourth, the attack does not require clock synchronization between the eavesdropper and the collider.

\subsubsection{Discussion on a simple attack detector}
\label{subsubsec:discussion}

A simple attack detection approach is to perform round-trip timing and then compare the measured round-trip time with a threshold. However, this approach has the following three shortcomings. First, it needs a downlink transmission for each uplink transmission, which doubles the communication overhead. LoRaWAN is mainly designed and optimized for uplinks. For instance, a LoRaWAN gateway can receive frames from multiple end devices simultaneously using different spreading factors, whereas it can send a single downlink frame only at a time. This is because Class A specification requires that any downlink transmission must be unicast, in response to a precedent uplink transmission. Thus, the round-trip timing approach matches poorly with the uplink-downlink asymmetry characteristic of LoRaWAN. Second, with this simple attack detection approach, it is the end device detecting the attack after receiving the downlink acknowledgement. The end device needs to inform the gateway using another uplink frame that is also subject to malicious collision. Third, as the attacks are rare (but critical) events, continually using downlink acknowledgements to preclude the threat is a low cost-effective solution. In summary, this simple round-trip timing countermeasure is inefficient and error-prone.

\subsection{Timing of Malicious Frame Collision}
\label{subsubsec:experiments}

\begin{figure}
  \centering
  \includegraphics[width=.7\columnwidth]{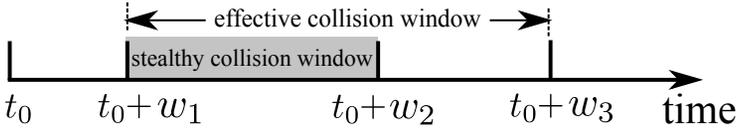}
  \caption{Collision attack time window.}
  \label{fig:time_window}
\end{figure}

In this section, we study the timing of effective malicious frame collision. When investigating the geographic area affected by the attack, the ratio between the powers of the victim signal and the collision signal also needs to be considered. \sect\ref{subsec:attack-surface} will jointly consider the collision timing and the signal power ratio. 
We set up two SX1276-based LoRa nodes as the transmitter and the receiver, which are separated by about $5\,\text{m}$. We use a third LoRa node as the collider against the receiver. The distance between the collider and the receiver is about $1\,\text{m}$. 
Although the quantified results obtained based on SX1276 are chip specific, the qualitative results (i.e., the trend) are consistent with the general understanding on wireless demodulation. Thus, the qualitative results provide general insights and implications. The gateway-class iC880A LoRaWAN concentrator and an open-source LoRa demodulator that we use in \sect\ref{subsec:attack-surface} also exhibit similar trend. In practice, the adversary may conduct experiments similar to those presented below to obtain the required attack timing once they know the model of the victim LoRa chip.

From our experiments, there are three critical time windows (denoted by $w_1$, $w_2$, and $w_3$) after the onset time of the victim transmission (denoted by $t_0$). These time windows are illustrated in Fig.~\ref{fig:time_window}. If the onset time of the collision frame is in $[t_0, t_0 + w_1]$, the receiver most likely receives the collision frame only; if it is in $[t_0 + w_1, t_0 + w_2]$, the receiver receives neither frame and raises no alerts; if it is in $[t_0 + w_2, t_0 + w_3]$, the receiver reports ``bad frame'' and yields no frame content; if it is after $t_0 + w_3$, the receiver can receive both frames sequentially. Therefore, the time window $[t_0 + w_1, t_0 + w_2]$ is called {\em stealthy collision window} and the $[t_0 + w_1, t_0 + w_3]$ is called {\em effective collision window}. Note that we view the ``bad frame'' situation as effective attack, because the receiver cannot differentiate malicious and normal collisions based on the warning message.

Our experiments measure $w_1$, $w_2$, and $w_3$ under a wide range of settings including spreading factor and the payload size. Table~\ref{table:attack-window} summarizes the results. From the results for $w_1$, the collision should start after the 5th chirp of the victim frame transmission. Explanation is as follows. (Note that as the demodulation mechanism of used SX1276 is proprietary and not publicly available, our explanations in this section are based on general understanding on wireless demodulation.) First, the receiver has not locked the victim frame's preamble until the 6th chirp. If the collision starts before the 5th chirp of the victim frame, the receiver will re-lock the collision frame's preamble with higher signal strength, resulting in reception of the collision frame. Second, the receiver locks the victim frame's preamble from the 6th chirp and simply drops any received radio data without reporting any error if any of the last three chirps (i.e., the 6th, 7th, and 8th chirps) of the preamble and/or the frame header are corrupted. For the latter case of frame header corruption, the radio chip cannot determine whether itself is the intended recipient and hence drops the received data. Thus, the collision should start after the 5th chirp of the victim frame.

We can also see that $w_2$ increases exponentially with the spreading factor. This is because: i) the total time for transmitting the preamble and frame header increases exponentially with the spreading factor; ii) corruption of the payload after the frame header leads to integrity check error and the ``bad frame'' message. The $w_3$ is roughly the time for transmitting the victim frame. Thus, if the collision onset time is after $t_0 + w_3$, both the victim and collision frames can be received.

\begin{table}
  \caption{Collision time windows for SX1276.}
  \label{table:attack-window}
  \centering
  \begin{tabular}{ccccccc}
    \toprule
    Spreading & Chirp & Preamble & Payload & $w_1$ & $w_2$ & $w_3$ \\
    factor $S$ & time & time & (byte) & & & \\
    \midrule
     & & & 10 & 5 & 28 & 141 \\
    7 & 1.024 & 8.2 & 20 & 5 & 38 & 156 \\
     & & & 30 & 6 & 41 & 165 \\
     & & & 40 & 6 & 54 & 178 \\
    \midrule
    7 & 1.024 & 8.2 & & 6 & 41 & 165 \\
    8 & 2.048 & 16.4 & 30 & 10 & 82 & 208 \\
    9 & 4.096 & 32.8 & & 22 & 156 & 274 \\
    \bottomrule
    \multicolumn{7}{l}{* Unit for chirp time, preamble time, $w_1$, $w_2$, $w_3$ is millisecond.} \\
  \end{tabular}
\end{table}

The above experiments show that, there is a time window of more than $20\,\text{ms}$ for the collision to corrupt the preamble partially and the frame header such that the victim simply drops the received data and raises no alerts. Collision starting in this window is stealthy. There is also an effective attack window of more than $100\,\text{ms}$. It is not difficult to satisfy such timing requirements using commodity radio devices.

\subsection{Size of Vulnerable Area}
\label{subsec:attack-surface}

In this section, through simulations and extensive experiments in a campus, we show that by setting up a collider and an eavesdropper at fixed locations, the frame delay attack can affect many end devices in a geographic area. The simulations based on realistic measurements with an open-source LoRa demodulator and a path loss model \cite{demetri2019automated} provide insights into understanding the vulnerable area. The experiments in the campus further capture other affecting factors such as terrain and signal blockage from buildings. In this section, the {\em core vulnerable area} refers to the geographic area in which the end devices are subject to stealthy collision and successful eavesdropping; the {\em vulnerable area} additionally includes the area in which the end devices are subject to the collision causing ``bad frame'' reports and successful eavesdropping.

\begin{figure}
  \centering
  \begin{minipage}[t]{.48\textwidth}
    \includegraphics{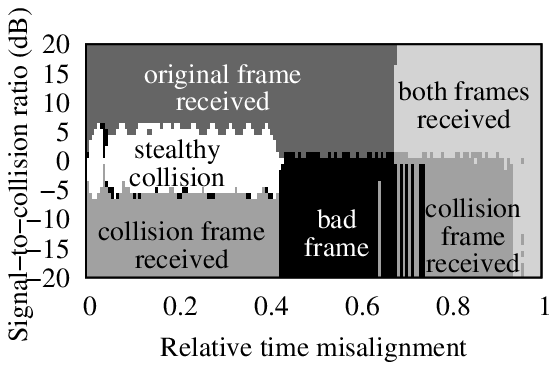}
    \caption{Result of \texttt{gr-lora}'s demodulation under collision with different signal-to-collision ratios and relative time misalignments.}
    \label{fig:diff_time_diff_power}
  \end{minipage}
  \hspace{0.5em}
  \begin{minipage}[t]{.48\textwidth}
    \includegraphics{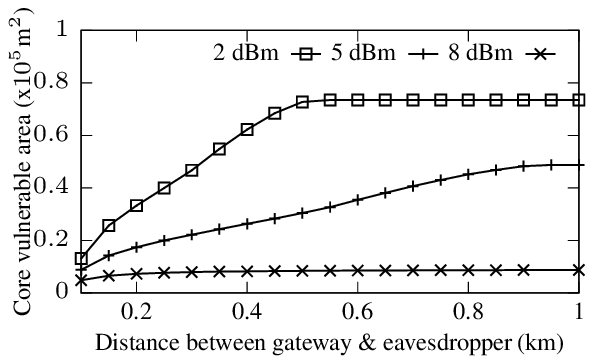}
    \caption{Core vulnerable area vs. distance between gateway and eavesdropper under various collision powers.}
    \label{fig:dist_area}
  \end{minipage}
\end{figure}

\begin{figure}
  \centering
    \includegraphics{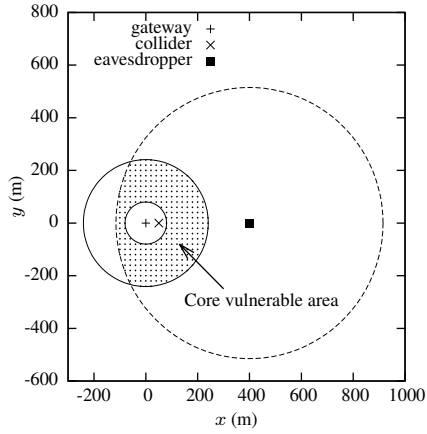}
    \caption{The core vulnerable area (i.e., the shaded area) with gateway at $(0,0)$, collider at $(50,0)$, and eavesdropper at $(400,0)$. Collider's and victim end device's transmitting powers are $2\,\text{dBm}$ and $14\,\text{dBm}$, respectively. End devices in the ring centered at $(0,0)$ are subject to stealthy collision; end devices in the dashed circle are subject to successful eavesdropping.}
    \label{fig:ideal_attack_surface}
\end{figure}

\begin{figure}
    \centering
    \includegraphics[width=0.9\textwidth]{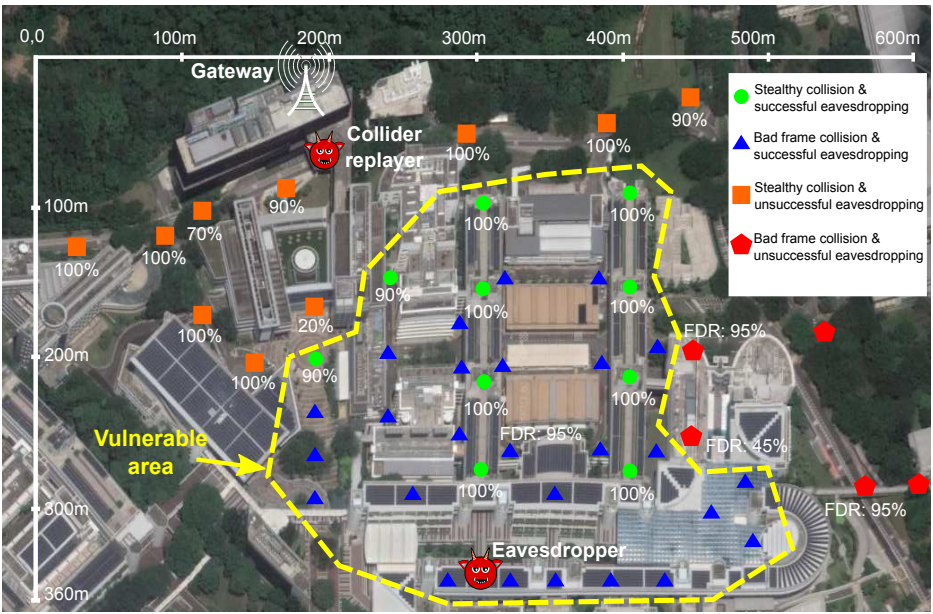}
    \caption{Vulnerable area of a campus LoRaWAN. A gateway and a USRP-based eavesdropper are deployed on the rooftops of two buildings. A collider is deployed on an overhead bridge. We carry an end device to each of the marked locations and conduct an attack experiment. The four point shapes represent four types of attack outcomes. (Satellite image credit: Google Map)}
    \label{fig:real_experiment_detailed}
\end{figure}

\subsubsection{Simulations}
\label{subsubsec:simulation}

To study the vulnerable area, we need to consider the signal path loss and the ratio between the powers of the victim signal and the collision signal at the receiver. We call this ratio {\em signal-to-collision ratio} (SCR). To characterize attack timing, we define {\em relative time misalignment} (RTM) as $\frac{\text{collision time lag}}{\text{frame time}}$, where the collision time lag is the time lag of the collision onset from the victim signal onset.
In our simulation, the victim and collision frames have identical length but different payload contents. 
We generate the $I$ and $Q$ waveforms of these two frames using LoRa signal model.
We superimpose the two frames' signals to simulate collision. Moreover, we scale the amplitudes of the two signals and time-misalign them to create certain SCR and RTM. The sum signal is processed using an open-source LoRa demodulator \texttt{gr-lora} \cite{grlora}.
Fig.~\ref{fig:diff_time_diff_power} shows the demodulation results under various SCR and RTM settings.
We can see that if RTM is less than 0.4 and SCR at the gateway is within $[-6\,\text{dB}, 6\,\text{dB}]$, the collision is stealthy. The eavesdropped frame can be demodulated if SCR at the eavesdropper is greater than $6\,\text{dB}$.

We adopt a LoRa signal path loss model for urban areas proposed in \cite{demetri2019automated} based on real measurements. The details of the model can be found in \cite{demetri2019automated}.
The frame delay attack is successful if the attacker can control RTM below 0.4 and satisfy the following two conditions:
\begin{align}
    -6\,\text{dB} &\leq P_{v} - L_{v,g} - (P_{c} - L_{c,g}) \leq 6\,\text{dB}, \label{eq:stealthy-condition} \\
    6\,\text{dB} &\leq P_{v} - L_{v,e} - (P_{c} - L_{c,e}), \label{eq:eavesdrop-condition}
    \vspace{-1em}
\end{align}
where the subscripts $v$, $g$, $c$, and $e$ respectively denote the victim end device, the gateway, the collider, and the eavesdropper; $P_x$ denotes the transmitting power of device $x$; $L_{x,y}$ denotes the path loss from device $x$ to $y$. Eq.~(\ref{eq:stealthy-condition}) is the condition for stealthy collision; Eq.~(\ref{eq:eavesdrop-condition}) is the condition for successful eavesdropping. The SCR thresholds of $6\,\text{dB}$ and $-6\,\text{dB}$ in Eqs.~(\ref{eq:stealthy-condition}) and (\ref{eq:eavesdrop-condition}) are from Fig.~\ref{fig:diff_time_diff_power}. Note that our modeling of successful eavesdropping in Eq.~(\ref{eq:eavesdrop-condition}) only considers the case that the signal from the collider at the eavesdropper has a power much higher than the noise floor, so that we can ignore the impact of noise on the eavesdropping.

Fig.~\ref{fig:ideal_attack_surface} shows an example of the areas defined by Eqs.~(\ref{eq:stealthy-condition}) and (\ref{eq:eavesdrop-condition}). The collider's and end device's transmitting powers are $2\,\text{dBm}$ and $14\,\text{dBm}$. The gateway's altitude is $25\,\text{m}$; the collider, eavesdropper, and end devices have an identical altitude of $0\,\text{m}$. As shown in Fig.~\ref{fig:ideal_attack_surface}, the ring centered at the gateway is defined by Eq.~(\ref{eq:stealthy-condition}); the disk area in the dashed circle is defined by Eq.~(\ref{eq:eavesdrop-condition}). Thus, the overlap between the ring and the disk is the core vulnerable area, which is $62,246\,\text{m}^2$.
Then, we vary the distance between the gateway and the eavesdropper (denoted by $d_{ge}$) and the $P_c$ setting. Fig.~\ref{fig:dist_area} shows the resulting core vulnerable area. We can see that the core vulnerable area in general increases with $d_{ge}$ and becomes flat after $d_{ge}$ exceeds a certain value. Moreover, among the three $P_c$ settings (i.e., 2, 5, and 8 dBm), $P_c = 2\,\text{dBm}$ gives larger core vulnerable areas. Reason of the above two observations is that the eavesdropper can achieve a larger eavesdropping area due to the weaker collision signal received by the eavesdropper. The core vulnerable area saturates because the eavesdropping area in the dashed circle illustrated in Fig.~\ref{fig:ideal_attack_surface} covers the entire ring area when $d_{ge}$ exceeds a certain value. Note that when $d_{ge}$ is very large, the noise power dominates and the core vulnerable area shrinks to zero.

The above simulation results suggest that the location of the gateway is the key information that the adversary needs to obtain. Based on that, the adversary can plan the placement of the collider and eavesdropper to affect a large geographic area. For the LoRaWANs adopting multiple gateways, the adversary can place a collider close to each of the gateways. In practice, the locations of the gateways can be obtained by the adversary in various ways (e.g., social engineering) and should not be relied on for the security of the system.

\subsubsection{Experiments in a campus LoRaWAN}
\label{subsubsec:exp_campus}
We conduct a set of experiments in an existing campus LoRaWAN to investigate the vulnerable area in real environments. The LoRaWAN's gateway covers the area shown in Fig.~\ref{fig:real_experiment_detailed} that has a number of multistory buildings.
The gateway, which consists of an iC880a LoRaWAN concentrator board, a Raspberry Pi, and a high-gain antenna, is located on the rooftop of a building. Both the collider and the eavesdropper consist of a laptop computer and a USRP N210 each. The collider is placed on an overhead bridge attached to the gateway's building. The horizontal distance between the gateway and the collider is about $50\,\text{m}$. The eavesdropper is placed on the rooftop of another building that is about $320\,\text{m}$ from the gateway's building. We carry an SX1276-based LoRaWAN end device to each of the locations marked in Fig.~\ref{fig:real_experiment_detailed}, measure the frame delivery ratio (FDR), and perform an attack experiment. The measured FDRs at all the visited locations are 100\%, except the four locations labeled with non-100\% FDRs. Thus, the gateway can cover the accessible area shown in Fig.~\ref{fig:real_experiment_detailed}.

In each attack experiment, the end device's and the collider's transmitting powers are $14\,\text{dBm}$ and $8\,\text{dBm}$, respectively. All malicious collisions are effective. The outcomes can be classified into four categories, which are the combinations of the collision results (stealthy collision or ``bad frame'') and eavesdropping results (successful or unsuccessful). In Fig.~\ref{fig:real_experiment_detailed}, we use four point shapes to represent the four attack outcomes. The percentage below a location is the ratio of stealthy collisions. We can see that, at most locations close to the gateway and collider, the malicious collisions are stealthy. At the locations in the bottom most part of Fig.~\ref{fig:real_experiment_detailed}, the collisions cause gateway's bad frame reports. There is a transit region in the middle of Fig.~\ref{fig:real_experiment_detailed}, in which the collision outcomes are mixed. Note that the visited locations shown in Fig.~\ref{fig:real_experiment_detailed} are on the rooftops, in semi-outdoor corridors, or in indoor environments. The indoor/outdoor condition may affect the collision outcome type. At the locations in the area enclosed by the dashed polygon, the gateway can decode the frame that is recorded by the eavesdropper and then replayed by the collider, suggesting that the eavesdropping is successful. Thus, this area is the vulnerable area caused by the attack setup, which is about $50,000\,\text{m}^2$.

Note that the demodulation mechanism of the iC880a concentrator is proprietary and can be different from the open-source LoRa demodulator we used in \sect\ref{subsubsec:simulation}. The actual signal propagation behaviors in the campus LoRaWAN can be much more complex than the model used in \sect\ref{subsec:attack-surface}. However, the simulation result (Fig.~\ref{fig:ideal_attack_surface}) and real experiment result (Fig.~\ref{fig:real_experiment_detailed}) show similar patterns, i.e., the eavesdropping area is around the eavesdropper and the core vulnerable area is a belt region between the gateway and the eavesdropper. Thus, our modeling and simulations in \sect\ref{subsec:attack-surface} provide useful understanding on the LoRaWAN vulnerability.

\section{$\mathsf{LoRaTS}$ Gateway}
\label{sec:lorasync}

As shown in \sect\ref{sec:security}, a fixed setup of a collider and an eavesdropper can subvert the sync-free timestamping for many end devices in a large geographic area. This section presents the $\mathsf{LoRaTS}$ gateway that supports the bandwidth-efficient sync-free timestamping as an advantage throughout the network lifetime and develops awareness of the frame delay attack.

\subsection{$\mathsf{LoRaTS}$ Gateway Hardware}
\label{subsec:hardware}

\begin{figure}
  \begin{minipage}[t]{.45\linewidth}
    \includegraphics[width=\textwidth]{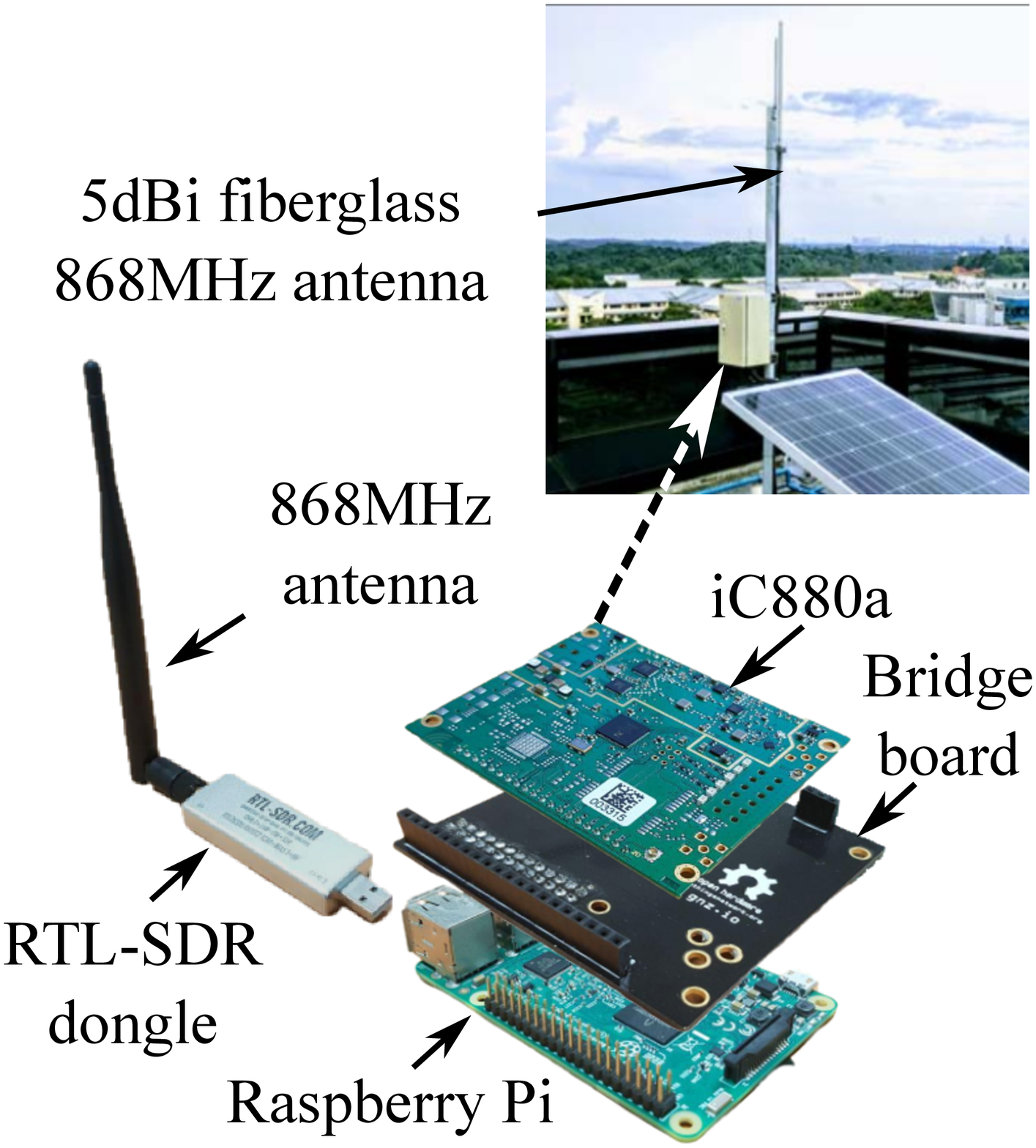}
    \caption{$\mathsf{LoRaTS}$ hardware prototype consisting of Raspberry Pi, iC880a concentrator, bridge board, RTL-SDR USB dongle.}
    \label{fig:softlora_gateway}
  \end{minipage}
  \hfill
  \begin{minipage}[t]{.45\linewidth}
    \includegraphics[width=\textwidth]{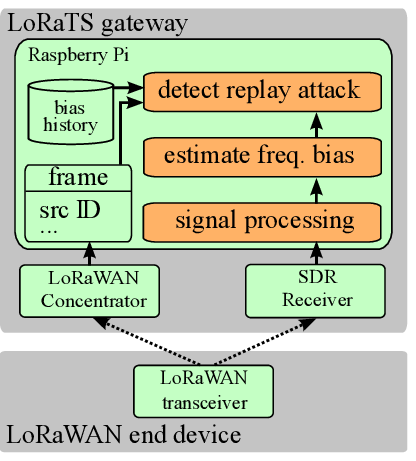}
    \caption{$\mathsf{LoRaTS}$ software. Bottom part is end device; upper part is gateway; solid arrows are local data flows; dashed arrows are transmissions.}
    \label{fig:lorawan_arch}
  \end{minipage}
\end{figure}

To detect the attack, we integrate an SDR receiver with a LoRaWAN gateway to monitor the physical layer. Various cheap (US\$25 only \cite{dongle-amazon}) and low-power SDR receivers are available now. In this paper, we use RTL-SDR USB dongles based on the RTL2832U chipset \cite{rtl-sdr}, which were originally designed to be DVB-T TV tuners. The RTL-SDR supports continuous tuning in the range of $[24, 1766]$ MHz, which covers the LoRaWAN bands.
It can operate at $2.4\,\text{Msps}$ reliably for extended time periods. Thus, the sampling resolution is $1/2.4\,\text{Msps} = 0.42\,\mu\text{s}$. Our research is conducted based on a $\mathsf{LoRaTS}$ hardware prototype that integrates a Raspberry Pi, an iC880a LoRaWAN concentrator, and an RTL-SDR USB dongle. Fig.~\ref{fig:softlora_gateway} shows the prototype.
An $868\,\text{MHz}$ antenna is used with the RTL-SDR to improve signal reception. 

The SDR receiver is used to capture the radio signal over a time duration of the first two preamble chirps of an uplink frame. 
The first sampled chirp is used to extract an accurate timestamp (cf.~\sect\ref{subsec:design}), whereas the second sampled chirp is used to extract the FB of the transmitter (cf.~\sect\ref{sec:fingerprint}). 
The accurate timestamp is a prerequisite of the FB estimation. 
As only two chirps' radio waveform is analyzed, the Raspberry Pi suffices for performing the computation.
Instead of using RTL-SDR, a full-fledged SDR transceiver (e.g., USRP) can be used to design a customized gateway with physical layer access. However, this design loses the factory-optimized hardware-speed LoRa demodulation built in the iC880a concentrator. Moreover, full-fledged SDR transceivers are often 10x more expensive than $\mathsf{LoRaTS}$. The low-cost, low-power, listen-only RTL-SDR suffices for developing the attack detector.

\subsection{$\mathsf{LoRaTS}$ Gateway Software}
\label{subsec:application}
The upper part of Fig.~\ref{fig:lorawan_arch} illustrates the software architecture of $\mathsf{LoRaTS}$ to detect the attack. It is based on the results in the subsequent sections of this paper.
The uplink transmission from the end device is captured by both the gateway's LoRaWAN concentrator and the SDR receiver. 
The LoRaWAN concentrator demodulates the received radio signal and passes the frame content to the Raspberry Pi. 
Signal processing algorithms are applied on the LoRa signal after down-conversion by the SDR receiver to determine precisely the arrival time of the uplink frame, estimate the transmitter's FB, and detect whether the current frame is a replayed one. 
The replay detection is by checking whether the estimated FB is consistent with the historical FBs associated with the transmitter ID contained in the current frame. 
Thus, the gateway is aware of the attack and can take necessary actions. 
Note that $\mathsf{LoRaTS}$ uses the SDR receiver to obtain FBs, rather than to decode the frame.

The LoRa communication is processed by the LoRa concentration board (i.e., iC880A), which is standalone hardware. The Raspberry Pi runs the concentration board's driver and forwards the packet data produced by the driver to our $\mathsf{LoRaTS}$ software stack.
Thus, the $\mathsf{LoRaTS}$ software stack and the driver-forwarder pipeline concurrently run on the Raspberry Pi.
As long as the computing resources of the Raspberry Pi are not exhausted, the $\mathsf{LoRaTS}$ software stack's computation and the communication will not have unhealthy contention. Note that the computing resources of the Raspberry Pi are sufficient to avoid the exhaustion.

\subsection{Signal Modeling and Uplink Frame Arrival Timestamping}
\label{subsec:design}

In this section, we present the modeling of the LoRa signal reception and our approach of detecting the onset time of the first preamble chirp. They form a basis for developing the FB estimation algorithms in \sect\ref{sec:fingerprint}.

\subsubsection{Derivation of $I$ and $Q$ Components of LoRa Signal}
\label{subsec:I-Q}

\begin{figure}
  \centering
  \includegraphics[width=0.9\linewidth]{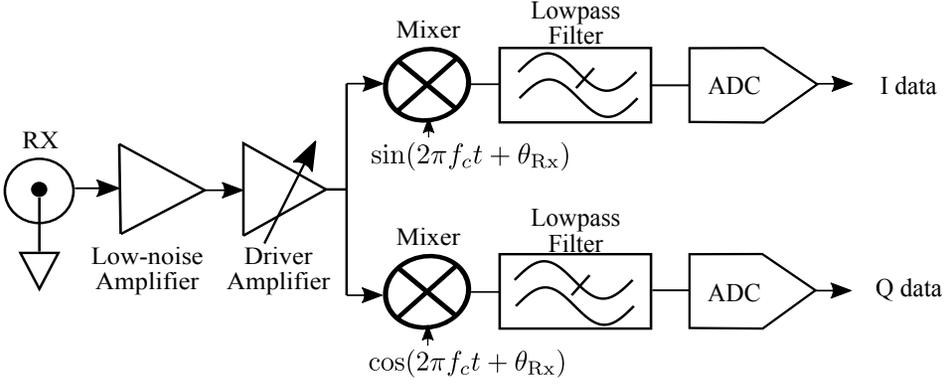}
  \caption{Analog signal processing in SDR receiver.}
  \label{fig:usrp}
\end{figure}

Fig.~\ref{fig:usrp} illustrates the essential analog signal processing steps of most SDR receivers to yield the in-phase ($I$) and quadrature ($Q$) components of the received radio signal.
The SDR receiver generates two unit-amplitude orthogonal carriers $\sin(2\pi f_ct + \theta_{\mathrm{Rx}})$ and $\cos(2 \pi f_c t + \theta_{\mathrm{Rx}})$, where $f_c$ is a specified frequency and $\theta_{\mathrm{Rx}}$ is the phase of the two self-generated carriers. The $f_c$ can be set to be the central frequency of the used LoRa channel. The $I$ and $Q$ components, denoted by $s_I(t)$ and $s_Q(t)$, are
\begin{align}
  s_I(t) = & s(t) \cdot \sin(2\pi f_c t + \theta_{\mathrm{Rx}}) \nonumber                                                                                                   \\
  =        & \frac{A(t)}{2} \left( \cos \left( 2\pi \int_0^tf(x)\dif x \!-\! 2\pi f_c t \!+\! \theta_{\mathrm{Tx}} \!-\! \theta_{\mathrm{Rx}} \right) \right. \label{eq:I0} \\
           & \left. - \cos \left( 2\pi \int_0^tf(x)\dif x + 2\pi f_c t + \theta_{\mathrm{Tx}} + \theta_{\mathrm{Rx}} \right) \right), \label{eq:high-freq-1}
\end{align}
\begin{align}
  s_Q(t) = & s(t) \cdot \cos \left( 2\pi f_c t + \theta_{\mathrm{Rx}} \right) \nonumber                                                                                         \\
  =        & \frac{A(t)}{2} \left( \sin \left(  2 \pi \int_0^t f(x) \dif x \!-\! 2\pi f_c t \!+\! \theta_{\mathrm{Tx}} \!-\! \theta_{\mathrm{Rx}} \right) \right. \label{eq:Q0} \\
           & \left. + \sin \left( 2\pi \int_0^t f(x) \dif x + 2\pi f_c t + \theta_{\mathrm{Tx}} + \theta_{\mathrm{Rx}} \right) \right),
  \label{eq:high-freq-2}
\end{align}
The high-frequency components in Eqs.~(\ref{eq:high-freq-1}) and (\ref{eq:high-freq-2}) are removed by the low-pass filters of the SDR receiver. Thus, the $I$ and $Q$ components after the filtering, denoted by $I(t)$ and $Q(t)$, are given by Eqs.~(\ref{eq:I0}) and (\ref{eq:Q0}). They can be rewritten as
\begin{align}
  I(t)      & = \frac{A(t)}{2} \cos \Theta(t), \quad Q(t) = \frac{A(t)}{2} \sin \Theta(t), \nonumber                                 \\
  \Theta(t) & = 2\pi \int_0^tf(x)\dif x - 2\pi f_c t + \theta, \quad \theta = \theta_{\mathrm{Tx}} - \theta_{\mathrm{Rx}}. \nonumber
\end{align}
The continuous-time $I(t)$ and $Q(t)$ are then sampled by the analog-to-digital converters (ADCs) to yield the $I$ and $Q$ data. For simplicity of exposition, the analysis in this paper is performed in the continuous-time domain.

\subsubsection{CSS reception using SDR receiver}
\label{subsec:preamble}
A chirp is a finite-time signal with time-varying frequency that sweeps the channel's bandwidth. Specifically, it can be expressed as $s(t) = A(t) \sin \Bigl( 2\pi \int_0^t f(x) \dif x + \theta_{\mathrm{Tx}} \Bigr)$, where $A(t)$ and $f(t)$ denote the instantaneous amplitude and frequency of the chirp at the time instant $t$, $\theta_{\mathrm{Tx}} \in [0, 2\pi)$ is the transmitter's phase that is usually unknown.

The SDR receiver generates two unit-amplitude orthogonal carriers $\sin(2\pi f_ct + \theta_{\mathrm{Rx}})$ and $\cos(2 \pi f_c t + \theta_{\mathrm{Rx}})$, where $f_c$ is the central frequency of the channel and $\theta_{\mathrm{Rx}}$ is the phase of the two self-generated carriers. After mixing the received signal with the two orthogonal carriers and applying low-pass filtering, which are standard operations of SDR, the SDR receiver yields the $I$ and $Q$ components that can be expressed as $I(t) = \frac{A(t)}{2} \cos \Theta(t)$ and $Q(t) = \frac{A(t)}{2} \sin \Theta(t)$, where the angle $\Theta(t) = 2\pi \int_0^tf(x)\dif x - 2\pi f_c t + \theta_{\mathrm{Tx}} - \theta_{\mathrm{Rx}}$. 

\begin{figure}
  \centering
  \subfigure[$I$ data]
  {
    \includegraphics[width=.40\linewidth]{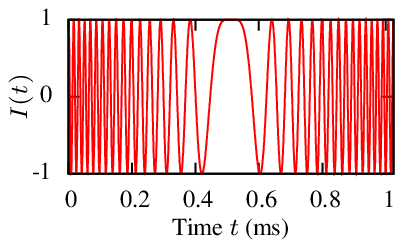}
  }
  \hspace{.02\linewidth}
  \subfigure[Spectrogram]
  {
    \includegraphics[width=.40\linewidth]{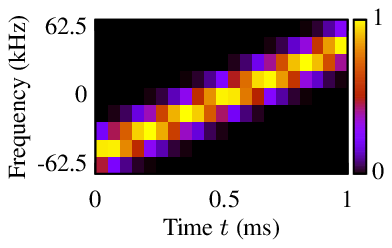}
  }
  \caption{$I$ data ($\theta_{\mathrm{Tx}}$=$\theta_{\mathrm{Rx}}$) and spectrogram of a preamble chirp.}
  \label{fig:up-chirp}
\end{figure}

A LoRaWAN uplink preamble consists of eight up chirps by default \cite{lorawan-spec}. For a preamble chirp, $f(t) = \frac{W^{2}}{2^{S}} \cdot t - \frac{W}{2} + f_c$ for $t \in \left[ 0, \frac{2^S}{W} \right]$, where $W$ is the channel bandwidth, $S \in \{6, 7, \ldots, 12\}$ is the {\em spreading factor}, and $\frac{2^S}{W}$ is the {\em chirp time}.
The $f(t)$ increases linearly from $(f_c - W/2)$ Hz to $(f_c + W/2)$ Hz over a chirp time. The angle of the preamble chirp can be derived as $\Theta(t) = \frac{\pi W^2}{2^S} t^2 - \pi W t + \theta_{\mathrm{Tx}} - \theta_{\mathrm{Rx}}$. In this paper, we use a channel with $f_c = 869.75\,\text{MHz}$ and $W = 125\,\text{kHz}$. Fig.~\ref{fig:up-chirp} shows the $I$ data and the spectrogram of an ideal preamble chirp. The parameters for generating Fig.~\ref{fig:up-chirp} are $A(t)=2$, $\theta_{\mathrm{Tx}}-\theta_{\mathrm{Rx}} = 0$, and $S=7$. Thus, the chirp time $\frac{2^S}{W}$ is $1.024\,\text{ms}$. To generate the spectrogram, we apply the short-time fast Fourier transform (FFT) with $2^S$-point Kaiser window and 16-point overlap between two neighbor windows. Thus, the spectrogram consists of 20 power spectral densities (PSDs) over the chirp time of $1.024\,\text{ms}$. 

\begin{figure}
    \subfigure[$I$ data with different phases.]
    {
      \includegraphics[width=.40\linewidth]{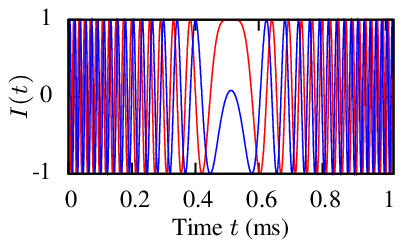}
      \label{fig:preamble_I_data_all}
    }
    \subfigure[Actual $I$ data of a preamble chirp with frequency bias.]
    {
      \centering 
      \includegraphics[width=.40\linewidth]{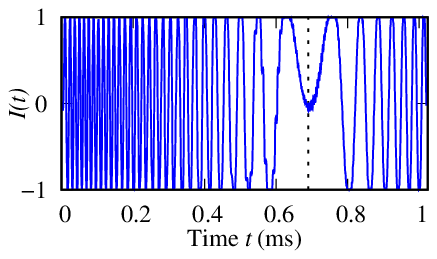}
      \label{fig:I_signal}
    }
    \caption{LoRa's $I$ signal waveform is affected by the initial phase and frequency bias (FB).}
    \label{fig:preamble_I_Q_data_all}
\end{figure}

\subsubsection{Preamble onset time detection}
\label{subsubsec:preamble-onset}
Detecting the onset time of the preamble is non-trivial. In this section, we discuss the matched filter approach and its inefficacy. Then, we present three other candidate methods.

\begin{figure}
  \centering
  \includegraphics[width=0.5\linewidth]{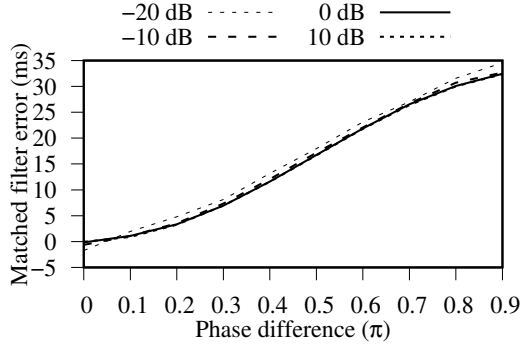}
  \caption{The matched filter error with different phase differences. Different lines represent different SNR.}
  \label{fig:phase_matched_error}
\end{figure}

Matched filter is a widely adopted symbol detection technique. 
Its principle is to slide a template signal over the incoming signal to detect the existence of the template signal's pattern in the incoming signal. Therefore, a basic assumption of the matched filter is that the I/Q signal of the received symbol has the same or similar pattern as the template signal. In the current context, this assumption is valid if the receiver is phase-locked to the transmitter (i.e., $\theta_{\mathrm{Rx}}=\theta_{\mathrm{Tx}}$). However, as LoRa adopts time-varying frequency, it is difficult for the SDR receiver to estimate the transmitter's phase $\theta_{\mathrm{Tx}}$.
As a result, the phase difference $\theta_{\mathrm{Tx}}-\theta_{\mathrm{Rx}}$, which is a critical factor affecting the pattern of $I(t)$ and $Q(t)$, will be random. Fig.~\ref{fig:preamble_I_data_all} shows the ideal $I(t)$ traces of the preamble chirp when the phase difference is $0$ and $\pi$, respectively. The waveform shapes are different. Thus, we cannot define a template signal to achieve efficient matched filtering. In addition, we conduct a numeric experiment to assess how the phase difference affects the matched filtering performance. Fig.~\ref{fig:phase_matched_error} presents the errors of the matched filter in determining the onset time of a preamble chirp ($S = 12$) using templates with different phase differences. Different curves correspond to different SNR. We can see that, compared with the noise level, the phase difference has a much more significant impact on the performance of the matched filter. The phase difference may result in up to tens of milliseconds error in onset time estimation. In contrast, the detectors presented in the rest of this section can achieve microseconds accuracy.

\begin{figure}
  \subfigure[Consecutive ratio]
  {
    \centering
      \includegraphics[width=.40\linewidth]{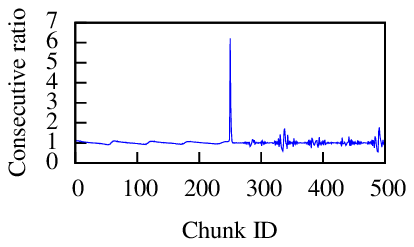}
      \label{fig:ccr_ratio}
  }
  \subfigure[ENV detector]
  {
    \centering
      \includegraphics[width=.40\linewidth]{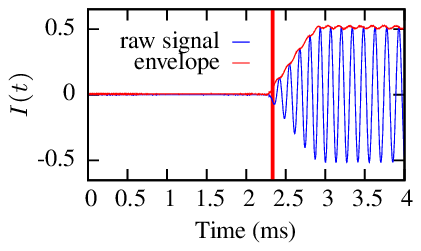}
      \label{fig:I_envelope}
  }
  \caption{The intermediate and final results of envelope detector.}
  \label{fig:envelope_detector}
\end{figure}

For $\mathsf{LoRaTS}$, we consider three parameter-less detectors:

\noindent {\bf Envelope (ENV) detector:} First, we apply the Hilbert transform to extract the amplitude envelope of the $I$ or $Q$ signal. Fig.~\ref{fig:I_envelope} shows the extracted amplitude envelope for $I$ data. 
We adopt the \textit{folding} technique \cite{zhou2010zifi,lovelace1969digital} to detect the signal onset time from the amplitude envelope. Specifically, we evenly divide the envelope into chunks of equal length (e.g., 200 samples). Then, we calculate the sum of the absolute values of the amplitudes of all samples in each chunk, which is referred to as \textit{trunk sum}. Lastly, we compute the ratio between the trunk sums of any two consecutive trucks to generate a ratio sequence. As shown in Fig.~\ref{fig:ccr_ratio}, the ratio sequence has a single peak. The detector yields the peak's time instant as the preamble onset time. The red vertical line in Fig.~\ref{fig:I_envelope} indicates the detected onset time.

\begin{figure}
  \centering
  \includegraphics[width=0.9\linewidth]{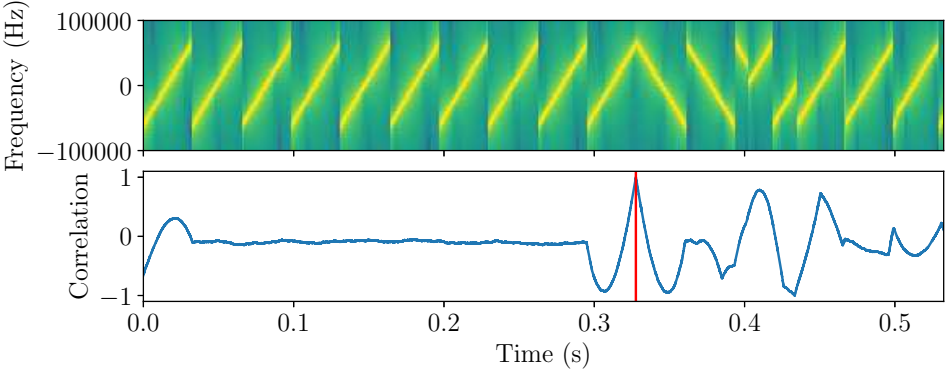}
  \caption{Correlation detector. The upper part shows the spectrogram of a LoRa frame. The bottom part shows the normalized correlation coefficient between the spectrogram and a locally generated hill peak pattern.}
  \label{fig:spec_auto_corr}
\end{figure}

\noindent {\bf Correlation (CORR) detector:} The Start Frame Delimiter (SFD) in a LoRa frame consists of two and a quarter down chirps. SFD is used by LoRa receiver for synchronization, because the junction of the up chirp before SFD and the first down chirp of SFD presents a salient hill peak as shown in the upper part of Fig.~\ref{fig:spec_auto_corr}. We can compute the correlation between the spectrograms of the received LoRa signal and a locally generated hill peak template. The maximum of the correlation trace gives the time instant of the hill peak, which can be used to infer the onset time of the LoRa frame. The bottom part of Fig.~\ref{fig:spec_auto_corr} shows the normalized correlation coefficient trace and the detected hill peak time represented by the red vertical line.

\noindent {\bf AIC detector:} The autoregressive Akaike Information Criterion (AIC) algorithm \cite{sleeman1999robust}
was originally developed to estimate the arrival time of seismic waves with an accuracy of a single sampling point.
As the $I$ and $Q$ signals are similar to the seismic waves \cite{liu2013volcanic}, the AIC is a promising solution for our problem.
It works as follows. For each point of the signal as an onset time candidate, two autoregressive models are constructed for the signal segments before and after the onset time candidate. The candidate that gives the largest dissimilarity between the two autoregressive models is yielded as the final result. From Fig.~\ref{fig:I_araic}, AIC can detect the onset time from the signal with a smooth start. From the results in \cite{sleeman1999robust}, AIC's detection results have a bias of 4 samples. With a sampling rate of $2.4\,\text{MHz}$, the bias is $\mathbb{E}[\epsilon]=\frac{4}{2.4\,\text{Msps}} = 1.67\,\mu\text{s}$ only, where $\epsilon$ represents onset time detection error.

\subsubsection{Evaluation}
\label{subsec:accuracy-evaluation}

\begin{figure}
  \centering
  \begin{minipage}[t]{.40\linewidth}
    \includegraphics[width=\linewidth]{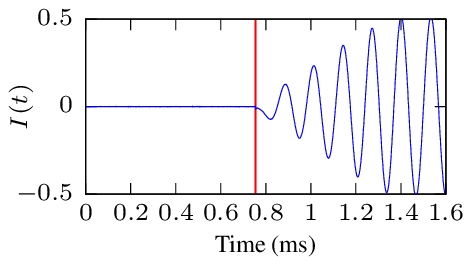}
    \caption{AIC detector}
    \label{fig:I_araic}
  \end{minipage}
  \hspace{.01\textwidth}
  \begin{minipage}[t]{.40\linewidth}
    \includegraphics[width=\textwidth]{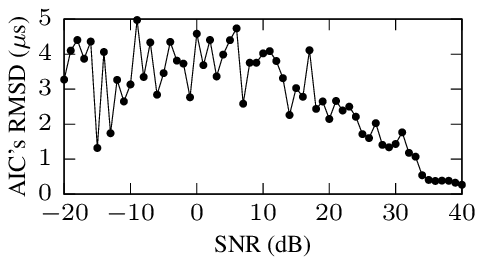}
    \caption{AIC's RMSD vs. SNR.}
    \label{fig:ar-aic-snr}
  \end{minipage}
\end{figure}

We conduct experiments to evaluate the performance of the three detectors presented above.
As AIC is nearly unbiased \cite{sleeman1999robust}, we primarily assess the root-mean-square deviation (RMSD), which characterizes the consistency of the detection results. Due to the difficulty in obtaining the ground truth of the preamble arrival time, we indirectly estimate the RMSD as follows. We place two $\mathsf{LoRaTS}$ nodes $A$ and $B$ close to each other such that the signal propagation time is near-zero. Node $A$ initiates a round-trip communication; each of them detects the onset times for both its transmitted and received signals, generating four onset times totally. Note that a $\mathsf{LoRaTS}$ node's SDR receiver can also capture the signal transmitted by the node's LoRa radio and detect the onset time. Denote by $\Delta$ the measured round-trip time based on the detected onset times; denote by $\epsilon_{X}^{Tx}$ and $\epsilon_{X}^{Rx}$ the node $X$'s unknown onset time detection errors for its transmitted and received signals. For the two close nodes $A$ and $B$, $\Delta = \epsilon_{A}^{Rx} - \epsilon_{A}^{Tx} + \epsilon_{B}^{Rx} - \epsilon_{B}^{Tx}$ and
$\mathrm{RMSD}(\epsilon) = \frac{1}{2} \mathrm{RMSD}(\Delta)$ if the errors are independent and identically distributed.
Measurements show that $\mathrm{RMSD}(\epsilon)$ is $1.21\,\mu\text{s}$, $0.64\,\mu\text{s}$, and $0.33\,\mu\text{s}$ for ENV, CORR, and AIC, respectively.
Thus, AIC achieves more consistent detection results.
Then, we evaluate the impact of random noises on AIC's $\mathrm{RMSD}(\epsilon)$. We artificially add zero-mean Gaussian noises to the collected high-SNR $I$ and $Q$ traces. Then, we apply AIC on the noise-added traces to detect the preamble onset time.
Fig.~\ref{fig:ar-aic-snr} shows the results. Note that the SNR range in Fig.~\ref{fig:ar-aic-snr} can cover realistic SNRs, e.g., $13\,\text{dB}$ to $-1\,\text{dB}$ in a multistory building (cf.~\sect\ref{sec:eval}). From Fig.~\ref{fig:ar-aic-snr}, the AIC's $\mathrm{RMSD}(\epsilon)$ is less than $5\,\mu\text{s}$ when the SNR is down to $-20\,\text{dB}$. Thus, AIC achieves robust onset time detection in the presence of strong noises. The rest of this paper uses AIC.


\section{Frame Delay Attack Detection}
\label{sec:fingerprint}
Internal oscillators for generating carriers generally have FBs due to manufacturing imperfection. This section develops algorithms for estimating LoRa transmitters' FBs based on LoRa's CSS modulation and use them to detect the frame delay attack. Note that the existing FB estimation algorithms developed for other radios cannot be ported to LoRa due to different modulation schemes. For instance, the FB estimation for OFDM \cite{yao2005blind} is apparently not applicable for LoRa CSS. As discussed later, LoRa demodulation's built-in FB estimation technique does not provide sufficient resolution. Thus,  highly accurate FB estimation for LoRa CSS is a non-trivial problem.

\subsection{FB Estimation}
\label{subsubsec:extraction}

This section describes algorithms for estimating the transmitter's FB based on an up chirp in the preamble.
First, we analyze the impact of the transmitter's and SDR receiver's FBs (denoted by $\delta_{\mathrm{Tx}}$ and $\delta_{\mathrm{Rx}}$) on the $I$ and $Q$ traces.
The up chirp's instantaneous frequency accounting for $\delta_{\mathrm{Tx}}$ is $f(t) = \frac{W^{2}}{2^{S}} \cdot t - \frac{W}{2} + f_c + \delta_{\mathrm{Tx}}$, $t \in \left[ 0, \frac{2^S}{W} \right]$. The two local unit-amplitude orthogonal carriers generated by the SDR receiver are $\sin(2 \pi (f_c + \delta_{\mathrm{Rx}})t + \theta_{\mathrm{Rx}})$ and $\cos(2 \pi (f_c + \delta_{\mathrm{Rx}})t + \theta_{\mathrm{Rx}})$. After mixing and low-pass filtering, the $I$ and $Q$ components of the received up chirp can be derived as $I(t) = \frac{A(t)}{2} \cos \Theta(t)$ and $Q(t) = \frac{A(t)}{2} \sin \Theta(t)$, where the angle $\Theta(t)$ is given by
\begin{equation}
  \Theta(t) = \frac{\pi W^2}{2^S} t^2 - \pi W t + 2\pi \delta t + \theta_{\mathrm{Tx}} - \theta_{\mathrm{Rx}}, \quad \delta = \delta_{\mathrm{Tx}} - \delta_{\mathrm{Rx}}. \label{eq:Theta-with-bias}
\end{equation}
When $\delta=0$, the axis of symmetry of $I(t)$ is located at the midpoint of the preamble chirp time. As shown in Fig.~\ref{fig:I_signal}, a negative $\delta$ causes a right shift of the axis of the symmetry in the time domain, whereas a positive $\delta$ causes a left shift.

For a certain SDR receiver, the FB estimation problem is to estimate $\delta$ from the captured $I$ and $Q$ traces. We do not need to estimate $\delta_{\mathrm{Tx}}$, because for a certain SDR receiver with a nearly fixed $\delta_{\mathrm{Rx}}$, a change in $\delta$ indicates a change in $\delta_{Tx}$ and a replay attack. In fact, FB estimation is a prerequisite of LoRa demodulation. Now, we discuss the incompetence of the LoRa demodulators' built-in FB estimation technique for attack detection. LoRa's CSS scheme evenly divides the whole channel bandwidth of $W\,\text{Hz}$ into $2^{S}$ bins, where $S$ is the spreading factor. The starting frequency of a bin corresponds to a symbol state. Since the preamble chirp linearly sweeps the channel bandwidth, its starting frequency can be viewed as the FB. LoRa demodulation firstly applies {\em dechirping} and then FFT to identify the preamble's and any data chirp's starting frequency bin indexes. The difference between the two indexes is the symbol state. As FFT achieves a resolution of $\frac{1}{x}\,\text{Hz}$ using $x$ seconds of data, the Fourier transform of a chirp with length of $\frac{2^S}{W}$ seconds has a frequency resolution of $\frac{W}{2^S}\,\text{Hz}$. This is also the resolution of the built-in FB estimation. Thus, for low spreading factor settings, the resolution may be poor. For instance, when $S=7$ and $W=125\,\text{kHz}$, the resolution is $976.56\,\text{Hz}$. As we will show in \sect\ref{subsec:countermeasures}, this near-$1\,\text{kHz}$ resolution is insufficient to detect attacks that introduce sub-$1\,\text{kHz}$ FBs. The colliding frame disentanglement approach Choir \cite{eletreby2017empowering} also uses the dechirping-FFT pipeline to analyze FB. Thus, it is subject to insufficient resolution. To achieve higher resolutions, this section presents two time-domain approaches, i.e., linear regression and least square, which are designed based on Eq.~(\ref{eq:Theta-with-bias}). 

The FB estimation problem is essentially a parameter estimation problem based on noisy data since the wireless channel is inevitably subjected to various noises. Different from the methodology of developing the optimal parameter estimation algorithm based on detailed assumptions about the channel, this paper uses the two widely adopted parameter estimation methods (i.e., linear regression and least squares) based on key insights obtained from the signal model. In particular, the least squares method is a rule-of-thumb approach to reduce the impact of noises.
We also conduct extensive evaluations to compare the two approaches' performance in the presence of various noise levels.

\subsubsection{Linear regression approach}
\label{subsubsec:linear-regression}
Eq.~(\ref{eq:Theta-with-bias}) can be rewritten as $\Theta(t) - \frac{\pi W^2}{2^S} t^2 + \pi W t = 2\pi \delta t + \theta$, which is a linear function of $t$ with $2\pi\delta$ as the slope. Thus, the slope can be estimated by linear regression based on the data pairs $(t, \Theta(t) - \frac{\pi W^2}{2^S} t^2 + \pi W t)$, where $t \in \left[ 0, \frac{2^S}{W} \right]$, $\Theta(t) = \mathrm{atan2}(Q(t), I(t)) + 2k\pi$, and $k \in \mathbb{Z}$ rectifies the multi-valued inverse tangent function $\mathrm{atan2}(\cdot, \cdot) \in (-\pi, \pi)$ to an unlimited value domain. 
The rectification is as follows. The $k$ is initialized to be $0$ when $t=0$. As $t$ increases, if $\mathrm{atan2}(Q(t), I(t))$ jumps from $-\pi$ to $\pi$, $k$ decreases by one; if $\mathrm{atan2}(Q(t), I(t))$ jumps from $\pi$ to $-\pi$, $k$ increases by one. Note that the traces $I(t)$ and $Q(t)$ where $t \in \left[0, \frac{2^S}{W} \right]$ are the segments of the captured $I$ and $Q$ signals starting from the preamble onset time detected by the AIC algorithm and lasting for a chirp time duration of $\frac{2^S}{W}$ seconds. 

Note that the $I(t)$ and $Q(t)$ are the $I$ and $Q$ data traces captured by the SDR receiver for a complete preamble chirp. The preamble onset time detected by AIC is used to segment the $I$ and $Q$ traces to chirps. From our measurements, the first preamble chirp, in general, has an increasing amplitude $A(t)$ after the onset time (as shown in Fig.~\ref{fig:I_araic}), which generates a negative impact on the linear regression accuracy. As the second preamble chirp has a more stable $A(t)$, we use the second chirp for the linear regression. 
As the linear regression approach has a closed-form formula to compute $\delta$, it has a complexity of $\mathcal{O}(1)$.

\subsubsection{Least squares approach}
\label{subsec:low-snr}
The LoRa signals can be very weak after long-distance propagation or barrier penetration. The LoRa's demodulation is designed to address low SNRs.
For SX1276, the minimum SNRs required for reliable demodulation with spreading factors of 7 to 12 are $-7.5\,\text{dB}$ to $-20\,\text{dB}$ \cite{sx1276-data}. We aim at extracting FB at such low SNRs. We solve a least squares problem:
\begin{align*}
  \argmin_{\theta_{\mathrm{Tx}} - \theta_{\mathrm{Rx}} \in [0, 2\pi), \delta} \sum_{t \in \left[ 0, 2^S/W \right]}  &\left(Q(t) - A \sin \Theta(t) \right)^2 + \left( I(t) - A \cos \Theta(t) \right)^2,
\end{align*}
where $Q(t)$ and $I(t)$ are the received $Q$ and $I$ traces; $\Theta(t)$ is given by Eq.~(\ref{eq:Theta-with-bias}); $A \sin \Theta(t)$ and $A \cos \Theta(t)$ are the noiseless $Q$ and $I$ templates. The above formulation requires that the $Q$ and $I$ templates have an identical and constant amplitude $A$. As the second preamble chirp can meet this requirement, we use it for FB estimation. The $A$ can be estimated as the square root of the difference between the average powers of the LoRa signal and the pure noise. 
We use a \texttt{scipy} implementation of the differential evolution algorithm \cite{storn1997differential} to solve the least squares problem. 
Raspberry Pi uses about 0.7 seconds to solve it. 
We use \textit{Memory Profile}~\cite{memory_profile}, a Python module for monitoring memory usage of Python programs, to profile our algorithm on a Raspberry Pi 3 Mod B. The memory usage is $77.191\,\text{MiB}$. The Raspberry Pi 3 Mod B has $1\,\text{GB}$ memory, which is sufficient for the computation.

\subsubsection{Performance comparison}
\label{subsubsec:detection_performance}
\begin{figure}
  \subfigure[Linear regression]
  {
    \centering
    \includegraphics[width=.40\linewidth]{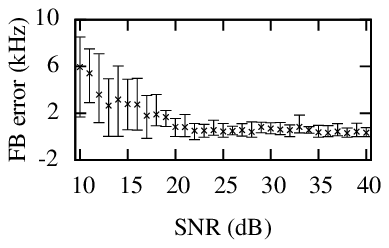}
    \label{fig:lr_estimation_error}
  }
  \subfigure[Least squares]
  {
    \centering
    \includegraphics[width=.40\linewidth]{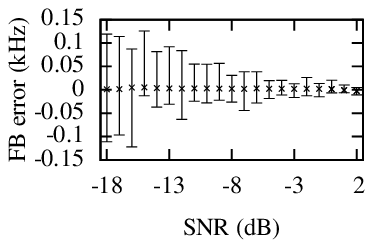}
    \label{fig:ls_estimation_error}
  }
  \caption{FB estimation errors vs. SNR.}
  \label{fig:fb-errors}
\end{figure}

We compare the FB estimation accuracy of the linear regression approach and the least squares approach.
Fig.~\ref{fig:fb-errors} shows the results. For each SNR setting, 20 LoRa $I$ and $Q$ traces with random FBs are generated using the signal model in Eq.~(\ref{eq:Theta-with-bias}). We also generate 20 noise traces; the magnitude of the noise is controlled to achieve the specified SNR. In Fig.~\ref{fig:fb-errors}, each error bar showing the 20\%- and 80\%-percentiles is from the 20 FB estimation results performed on the sum signals of the generated ideal LoRa signals and noise. From Fig.~\ref{fig:lr_estimation_error}, the linear regression approach can achieve low FB estimation errors when the SNR is very high (e.g., $40\,\text{dB}$). However, it performs poorly for low SNRs. This is caused by the susceptibility of the inverse tangent rectification to noises. Specifically, as the inverse tangent rectification is based on a heuristic to detect $\mathrm{atan2}$'s sudden transitions between $-\pi$ and $\pi$, large noises lead to false positive detection of the transitions.
Differently, the least squares approach maintains the FB estimation error within $120\,\text{Hz}$ (i.e., $0.14\,\text{ppm}$), when the SNR is down to $-18\,\text{dB}$. Thus, the rest of this paper adopts the noise-resilient least squares approach, though it is more compute-intensive.

\subsubsection{FB measurements for 16 end devices}
\label{subsubsec:fb_indoor}
\begin{figure}
  \centering
  \includegraphics[width=0.9\linewidth]{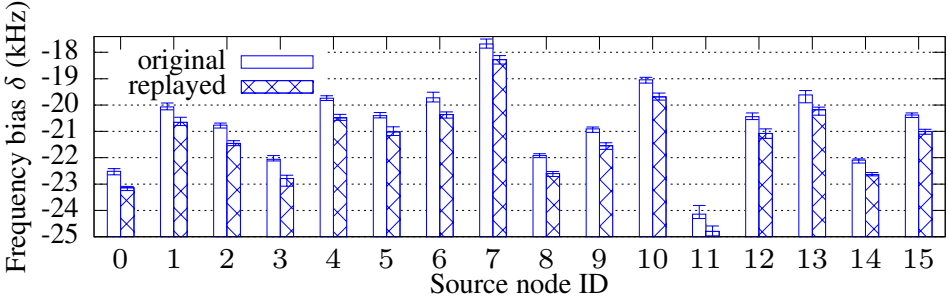}
  \caption{FBs estimated from the original LoRa signals from 16 end devices and those replayed by a USRP-based replayer. The error bar shows mean, minimum, and maximum of FBs in 20 frame transmissions.}
  \label{fig:frequency_bias}
\end{figure}
We use an RTL-SDR to estimate the FBs of 16 SX1276-based end devices. In each test for an end device, the distance between the end device and the RTL-SDR is about $5\,\text{m}$. The error bars labeled ``original'' in Fig.~\ref{fig:frequency_bias} show the results.
We can see that the FBs for a certain node are stable, and the nodes generally have different FBs. The absolute FBs are from $17\,\text{kHz}$ to $25\,\text{kHz}$, which are about $20\,\text{ppm}$ to $29\,\text{ppm}$ of the nominal central frequency of $869.75\,\text{MHz}$. Some nodes have similar FBs, e.g., Node 3, 8, and 14.
Note that the detection of the replay attack is based on the fact that the replayed transmission has a different FB. In other words, the attack detection does not require distinct FBs among different end devices. From Fig.~\ref{fig:frequency_bias}, we also observe that all nodes have negative FB measurements, which means that $\delta_{\mathrm{Tx}} < \delta_{\mathrm{Rx}}$, where $\delta_{\mathrm{Tx}}$ and $\delta_{\mathrm{Rx}}$ are the unknown FBs of the end device and the RTL-SDR. Note that as the RTL-SDR is a low-cost device, it may have a large FB causing the negative relative FB measurements.

\subsubsection{Impact of coding rate and bandwidth settings on FB estimation}
\label{subsubsec:crc_bw_settings}
\begin{figure}
  \centering
  \includegraphics[width=0.5\linewidth]{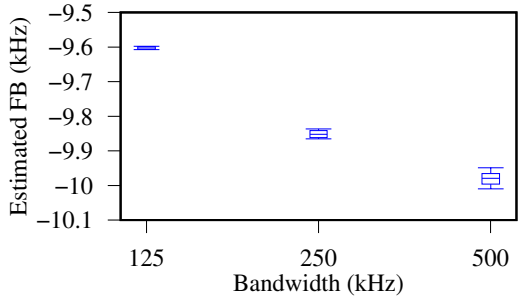}
  \caption{FBs estimated from an end device with different bandwidth settings. Each box plot shows minimum, maximum, median, 25\%, and 75\% percentiles.}
  \label{fig:diff_bw}
\end{figure}
In this paper, we use a single preamble chirp to estimate the FB. The coding rate, which is related to the forward error correction (FEC) for payload, is irrelevant to the preamble chirps. Therefore, the coding rate is irrelevant to our FB estimation method. 
Regarding signal bandwidth, as our FB estimation approach is developed based on the signal model with the bandwidth as a parameter, the approach is valid under any bandwidth setting.
We conduct an experiment to check the impact of different bandwidth settings on the FB estimation. We set $S = 12$, $f_c = 869.75\,\text{MHz}$, $\text{coding rate} = 4/5$ while changing the bandwidth $W$ to all possible settings, i.e., $125\,\text{kHz}$, $250\,\text{kHz}$, and $500\,\text{kHz}$. Fig.~\ref{fig:diff_bw} presents the FB estimates for an end device with different bandwidth settings. We can see that the estimation result varies slightly with the bandwidth setting. Thus, $\mathsf{LoRaTS}$ should re-profile the FB when an end device changes its bandwidth at run time.

\subsection{Replay Attack Detection}
\label{subsec:countermeasures}

The replayer also has an FB. The error bars labeled ``replayed'' in Fig.~\ref{fig:frequency_bias} show the FBs estimated from the LoRa signals received by the $\mathsf{LoRaTS}$'s SDR receiver when a USRP replays the radio waveform captured by itself in the experiments presented in \sect\ref{subsubsec:extraction}.
Compared with the results labeled ``original'', the FBs of the replayed transmissions are consistently lower. This is because the USRP has a negative FB.
The average additional FBs introduced by the replayer range from $-543$ to $-743\,\text{Hz}$, i.e., $0.62$ to $0.85\,\text{ppm}$  of the channel's central frequency. Thus, with the FB estimation accuracy of $0.14\,\text{ppm}$ achieved under low SNRs (cf.~\sect\ref{subsec:low-snr}), the additional FBs caused by the replay attack can be detected.

Based on the above observation, we describe an approach to detect the delayed replay. $\mathsf{LoRaTS}$ maintains a database of the FBs of the nodes with which it communicates. This database can be built offline or at run time using its SDR receiver in the absence of attacks. To address the end devices' time-varying radio frequency skews due to run-time conditions like temperature, $\mathsf{LoRaTS}$ can continuously update the database entries based on the FBs estimated from recent frames. To decide whether the current received frame is a replayed frame, the $\mathsf{LoRaTS}$ gateway checks whether the FB of the current received frame is within the acceptable FB range of the end device based on the database. This detection approach is applied after the $\mathsf{LoRaTS}$ gateway decodes the frame to obtain the end device ID. The FB estimated from a frame detected as a replayed one should not be used to update the database.

This detection mechanism forms the first line of defense against the frame delay attacks that introduce extra FBs. It gives awareness of the attack that is based on the logistics of collision and record-and-replay. Note that with the knowledge of our detector, the attackers may invest more resources and efforts to hide their radiometrics. Thus, our FB-based attack detector forces the attackers to hide their radiometrics with increased cost and technical barriers.\sect\ref{sec:zero-fb} will discuss the approach to eliminate the extra FBs and a further countermeasure to deal with the more crafted attack. 


\section{Zero-FB Attack and Countermeasure}
\label{sec:zero-fb}
\subsection{Implementation of Zero-FB Attack}

\begin{figure}
    \centering
    \includegraphics[width=.4\columnwidth]{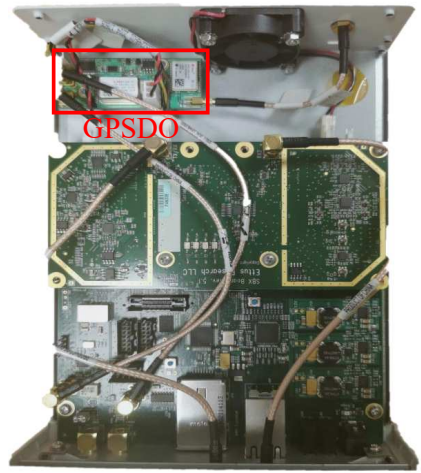}
    \caption{USRP N210 with GPSDO. The GPSDO module is marked with a red rectangle.}
    \label{fig:N210_GPSDO}
\end{figure}
  
To bypass the FB-based attack detector described in \sect\ref{subsec:countermeasures}, the adversary needs to precisely calibrate its eavesdropper and replayer to have FBs lower than the resolution of our FB estimation algorithm, e.g., $0.14\,\text{ppm}$ as shown in \sect\ref{subsec:low-snr}. We call this attack {\em zero-FB attack}. The zero-FB attack's calibration requires a highly accurate frequency source operating at the channel frequency, which is non-trivial. The GPSDO module of USRP provides a GPS-locked reference clock of $10\,\text{MHz}$ with 0.025 ppm accuracy \cite{gpsdo}. In this section, we investigate whether we can implement the zero-FB attack using GPSDO-equipped USRP. Specifically, as shown in Fig.~\ref{fig:N210_GPSDO}, we add a GPSDO module to the USRP-based eavesdropper and replayer, respectively. We use this improved attack apparatus to launch the frame delay attack. At the same time, we use $\mathsf{LoRaTS}$ to estimate the FBs of the original frame and the replayed frame for attack detection. Based on the experiment results, the additional FBs introduced by the attack are less than 0.035 ppm. The result is consistent with the accuracy level of the GPSDO's clock. Moreover, as this additional FB is below the resolution of our FB estimation algorithm (i.e., $0.14\,\text{ppm}$), our proposed FB-based attack detector is incapable of effectively detecting the attack. However, we stress that the attack setup is costly -- each GPSDO unit costs about US\$1,800 and the attack apparatus enhancement costs about US\$3,600 in total, since both the collider and the eavesdropper need the GPSDO module. Therefore, we use a US\$25 RTL-SDR to enforce the attacker to invest  US\$3,600 more in order to win the attack-defense chase.

\subsection{Countermeasure}
\label{subsec:zero_fb_countermeasure}
To detect the zero-FB attack, we propose a Pseudorandom Interval Hopping (PIH) scheme. The key idea of PIH is that, if the end device and the gateway achieve an agreement on the inter-frame intervals that are pseudorandom, the gateway can check the actual inter-frame intervals against the prior agreed intervals to detect the delays introduced by the attacker. To achieve the prior agreement, the gateway and the end device can establish a secret symmetric key by a key establishment protocol such as Diffie-Hellman's and use a pseudorandom generator seeded with the symmetric key to generate the inter-frame intervals. As the seed is confidential, the attacker can hardly predict the future pseudorandom inter-frame intervals based on the observed intervals. The end device will follow the pseudorandom sequence to regulate the transmission interval in transmitting the data frames. Now, we discuss several issues related to PIH.

\vspace{.5em}
\noindent {\bf Dealing with the waiting time at the end device.} PIH is an add-on to sync-free timestamping for enhancing security. It does not require clock synchronization between the end device and the gateway because the gateway only checks the transmission interval between two consecutive frames. When the end device generates a new data record, it needs to wait for the next scheduled transmission time. Therefore, the end device needs to inform the gateway of the waiting time for correct data timestamping. The details of dealing with this issue are as follows. A LoRaWAN end device records the times of interest  (e.g., the time instants when new sensor data records are acquired) in terms of its unsynchronized clock. Right before sending a number of data records using a frame, the device replaces the records' times of interest in its local clock with their elapsed times up to the present, form the frame, and transmit it immediately. We assume that the waiting time from the generation to the transmission of the data records is short to ensure limited local clock drift and limited bits to represent the elapsed times. For instance, to enforce an upper bound of $10\,\text{ms}$ clock drift under a drift rate of 40 ppm, the waiting time needs to be within 4.1 minutes. Thus, the sync-free synchronization can compensate for the arbitrary transmission interval introduced by the PIH. Note that the waiting time also needs to account for attack detection capability that is discussed shortly.

\vspace{.5em}
\noindent {\bf Setting of maximum inter-frame interval.}
Due to the end device's clock drift, the inter-frame interval may not exactly follow the prior agreed interval. To deal with this issue, the gateway can set a threshold to tolerate some deviations. If more deviation is allowed, the zero-FB attack can introduce some delay below the allowed deviation and remains undetected. If less deviation is allowed, the gateway may generate excessive false alarms, and the end device may need to use frequent synchronization sessions to calibrate its clocks, which introduces communication overhead. Thus, the threshold setting should be studied to achieve a good balance between security and efficiency under the PIH scheme. We provide an analysis as follows to guide the setting of the maximum inter-frame interval. The timing error of the end device, denoted by $E$, can be represented by $E = t \cdot r$, where $t$ and $r$ denote the maximum interval and clock drift rate, respectively. For example, if the $r$ is $40\,\text{ppm}$ (same setting as \sect\ref{subsec:sync-vs-free}) and the maximum interval $t$ is 30 minutes, the maximum timing error is $72\,\text{ms}$. Since we assume that the gateway has UTC, the allowed deviation can be determined based on the timing error of the end device. Reversely, we can determine the maximum interval based on a certain allowed deviation. For instance, if we set the allowed deviation to be $10\,\text{ms}$, the interval should be no greater than $\lceil\frac{E}{r}\rceil$ (i.e., $\lceil\frac{10\,\text{ms}}{40\,\text{ppm}}\rceil = 25\,\text{s}$). The pseudorandom generator needs to follow a uniform distribution, where the upper bound is the maximum interval. Moreover, the clock drift rate of the end device with Temperature Compensated Crystal Oscillator (TCXO) is low (e.g., $1\,\text{ppb}$). The parameter settings will be more flexible on these end devices with TCXOs. We note that the total clock drift of the end device over multiple uplink transmissions is irrelevant to the PIH approach that only checks the interval between any two consecutive uplink transmissions.

\vspace{.5em}
\noindent {\bf Transmission skipping, timeliness improvement, and frame losses.}
Sometimes the end device does not have data pending transmission at a scheduled transmission time or has data that is to be transmitted as soon as possible. To improve the efficiency and timeliness of PIH, we can divide the time into shorter time slices and allow skipping transmissions at the scheduled time instants. In this way, the real-time performance can be improved. In the case of a lossy link, the gateway can check the frame counter in the uplink frame and check the sum of these inter-frame intervals.
\section{Experiments}
\label{sec:eval}
\subsection{Experiments in a Multistory Building}
\label{sec:indoor-exp}
\begin{figure}
  \centering
  \includegraphics[width=0.9\linewidth]{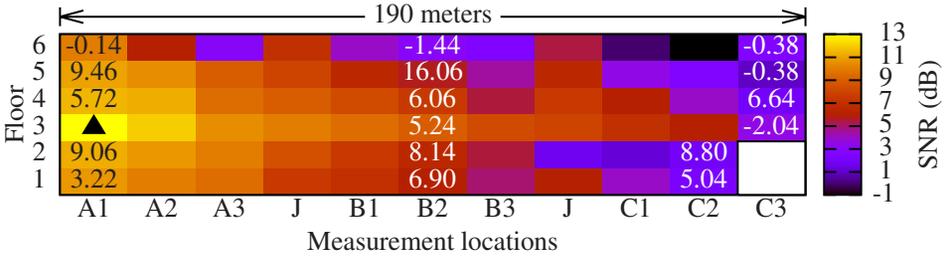}
  \caption{SNR survey in a building (lateral view) with 3 sections (A, B, C) and 2 junctions (J). The triangle represents the fixed node. The number in a cell is round-trip time measurement in $\mu$s excluding propagation time but containing onset time detection error when mobile node is in the cell.}
  \label{fig:snr-survey}
\end{figure}

LoRaWAN can be used for indoor applications, such as utility metering. We conduct a set of experiments to investigate the feasibility of attack and effectiveness of our attack detector in a concrete building with six floors. The building has three sections and two section junctions along its long dimension of 190 meters. Fig.~\ref{fig:snr-survey} illustrates a lateral view of the building. First, we survey the SNR inside the building to understand the signal attenuation. We deploy a fixed LoRaWAN transmitter in Section~A on the 3rd floor. Then, we carry an SDR receiver to different positions inside the building to measure the SNR. At each position, we first profile the noise power and then measure the total power when the fixed node transmits.
In each section, we measure three positions. The heat map in Fig.~\ref{fig:snr-survey} shows the SNR measurements. We can see that the SNR decays with the distance between the two nodes. The SNRs are from $-1\,\text{dB}$ to $13\,\text{dB}$.
Then, we conduct the following experiments. By default, we set $S = 12$, $f_c = 869.75\,\text{MHz}$, $W=125\,\text{kHz}$, $\text{coding rate} = 4/5$.

\vspace{0.2em}
\noindent {\bf Attack experiments:}
We deploy an iC880a-based gateway and an SX1276-based end device in Section A1 of the 3rd floor and Section C3 of the 6th floor, respectively.
The LoRa signals are significantly attenuated after passing through multiple building floors. If the end device adopts a spreading factor of 7, it cannot communicate with the gateway. A minimum spreading factor of 8 is needed for communications.
We deploy two USRP N210 stations as the eavesdropper and the collider, next to the end device and the gateway, respectively.
We set the transmitting power of the end device and the collider to be $14\,\text{dBm}$. 
The malicious collision is stealthy to the gateway; the eavesdropping is successful.
Thus, the frame delay attack can be launched in this building.

\vspace{0.2em}
\noindent {\bf Onset time detection:} We replace the iC880a gateway with our $\mathsf{LoRaTS}$ gateway and move the end device in the building.
The number shown in a cell of Fig.~\ref{fig:snr-survey} is the measured round-trip time $\Delta$ in $\mu$s excluding the propagation delay when the end device is at the corresponding location. Note that the propagation delay is calculated based on the estimated Euclidean distance between the gateway and the end device. As $\Delta$ contains an onset time detection error, it may become negative. Then, we compute the RMSD of all $\Delta$ measurements shown in Fig.~\ref{fig:snr-survey}. From the analysis in \sect\ref{subsec:accuracy-evaluation}, the AIC's average $\mathrm{RMSD}(\epsilon)$ in this building is $2.4\,\mu\text{s}$ only. This result is consistent with that in Fig.~\ref{fig:ar-aic-snr}.

\begin{figure}
  \centering
  \includegraphics[width=0.9\linewidth]{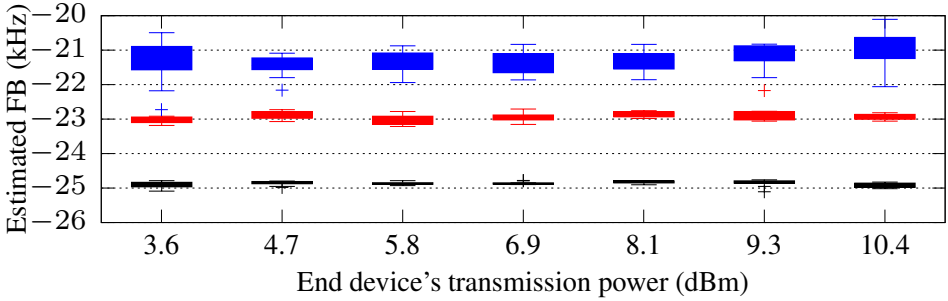}
  \caption{Estimated FB vs. transmitting power of the end device. Each box plot shows min, max, 25\% and 75\% percentiles. (1) Bottom row: end device to eavesdropper; (2) Mid row: end device to $\mathsf{LoRaTS}$ gateway; (3) Top row: replayer to $\mathsf{LoRaTS}$.}
  \label{fig:delta-power}
\end{figure}

\vspace{0.2em}
\noindent {\bf Impact of transmitting power on FB estimation:} Fig.~\ref{fig:delta-power} shows the estimated FBs versus the end device's transmitting power under different settings. 
The bottom row of black box plots is the FBs estimated by the eavesdropper when the end device transmits the uplink frame with different transmitting powers. 
The middle row of red box plots is the FBs estimated by the $\mathsf{LoRaTS}$ gateway in the absence of the frame collision and replay attacks. Thus, the FBs estimated by the eavesdropper and the $\mathsf{LoRaTS}$ gateway are different. 
This is because that as analyzed in \sect\ref{subsubsec:extraction}, the estimated FB $\delta$ contains the transmitter's and receiver's FBs $\delta_{\mathrm{Tx}}$ and $\delta_{\mathrm{Rx}}$. 
Note that the eavesdropper and the $\mathsf{LoRaTS}$ gateway, in general, have different FBs. From Fig.~\ref{fig:delta-power}, the end device's transmitting power has little impact on the FB estimation.

\vspace{0.2em}
\noindent {\bf Additional FB introduced by replayer:}
In Fig.~\ref{fig:delta-power}, the top row of blue box plots are the FBs estimated by the $\mathsf{LoRaTS}$ when the replayer replays the radio waveform recorded by the eavesdropper. 
When the end device adopts a higher transmitting power, the replayed signal also has higher power. By comparing the middle and the top rows, we can see that the replay attack introduces an additional FB of about $2\,\text{kHz}$, which is $2.3\,\text{ppm}$ of the LoRa channel's central frequency. Therefore, FB monitoring can easily detect the replay attack. Compared with the results in Fig.~\ref{fig:frequency_bias} showing additional FBs of $0.62$ to $0.85\,\text{ppm}$, the FBs in this set of experiments are higher. This is because we use two different USRPs as the eavesdropper and replayer; their FBs are superimposed.

\subsection{Temporal Stability of FB}
\label{subsec:temporal_stability}
\begin{figure}
  \subfigure[FB \& temperature.]
  {
    \centering
    \includegraphics[width=0.37\linewidth]{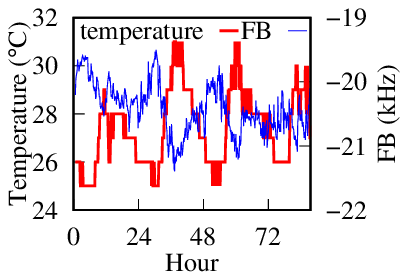}
    \label{fig:fb_temp}
  }
  \subfigure[FB variation CDF.]
  { 
    \centering
    \includegraphics[width=0.26\linewidth]{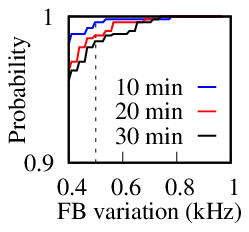}
    \label{fig:fb_cdf}
  }
  \subfigure[SX1262 FBs.]
  {
    \centering
    \includegraphics[width=0.26\linewidth]{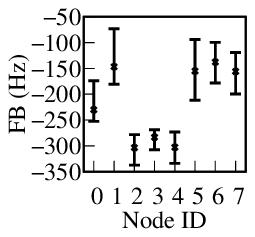}
    \label{fig:TCXO_fb}
  }
  \caption{(a), (b): Temporal stability of SX1276's FB over 87 hours. (c): The FBs of eight SX1262 LoRa chips with TCXOs.}
\end{figure}

FB can be affected by ambient conditions such as temperature. We continuously track the FB of an SX1276-based end device for 87 hours to study its temporal stability. We place the end device with a temperature sensor in a semi-outdoor corridor with time-varying temperature.
The end device transmits 10 frames every 10 minutes to the $\mathsf{LoRaTS}$ gateway, resulting 1,440 frames per day. Fig.~\ref{fig:fb_temp} shows the end device's temperature and FB traces.
The Pearson correlation between FB and temperature is $-0.78$.
Moreover, the FB has transient variations that can be caused by interference from other communication systems operating in neighbor frequency bands. As $\mathsf{LoRaTS}$ detects the attack based on the changes of FB, such transient variations may cause false alarms. Fig.~\ref{fig:fb_cdf} shows the CDFs of the maximum FB variation if the end device transmits a frame every 10, 20, and 30 minutes. If the attack detection threshold is $500\,\text{Hz}$ based on our previous measurements of the additional FB introduced by the attack, from the CDFs, the false alarm rate (i.e., the probability that the FB variation exceeds $500\,\text{Hz}$) is about 0.4\%, 1.3\%, and 1.7\% for the three frame interval settings.

The SX1276 used in this paper does not have TCXO. For LoRa radios with TCXO, the false alarm rate can be further reduced.
To verify this, we deploy eight SX1262-based end devices. SX1262 is the next-generation LoRa chip equipped with TCXO \cite{sx1262}. In Fig.~\ref{fig:TCXO_fb}, each error bar shows the 10\%- and 90\%-percentiles of 150 FB estimation results. We can see that the TCXO can significantly shrink the fluctuation. Specifically, from our measurements, the FB variations are no greater than $250\,\text{Hz}$. In contrast, from Fig.~\ref{fig:fb_temp}, without TCXO, the FB fluctuation range is up to $1\,\text{kHz}$. Thus, with a detection threshold of $500\,\text{Hz}$, the false alarm rate of our approach for a system based on SX1262 will be near-zero.

\subsection{Outdoor Experiments}
\label{subsec:outdoor_experiment}
\begin{figure}
  \subfigure[View parklot from rooftop]
  {
    \includegraphics[height=3cm, width=0.29\linewidth]{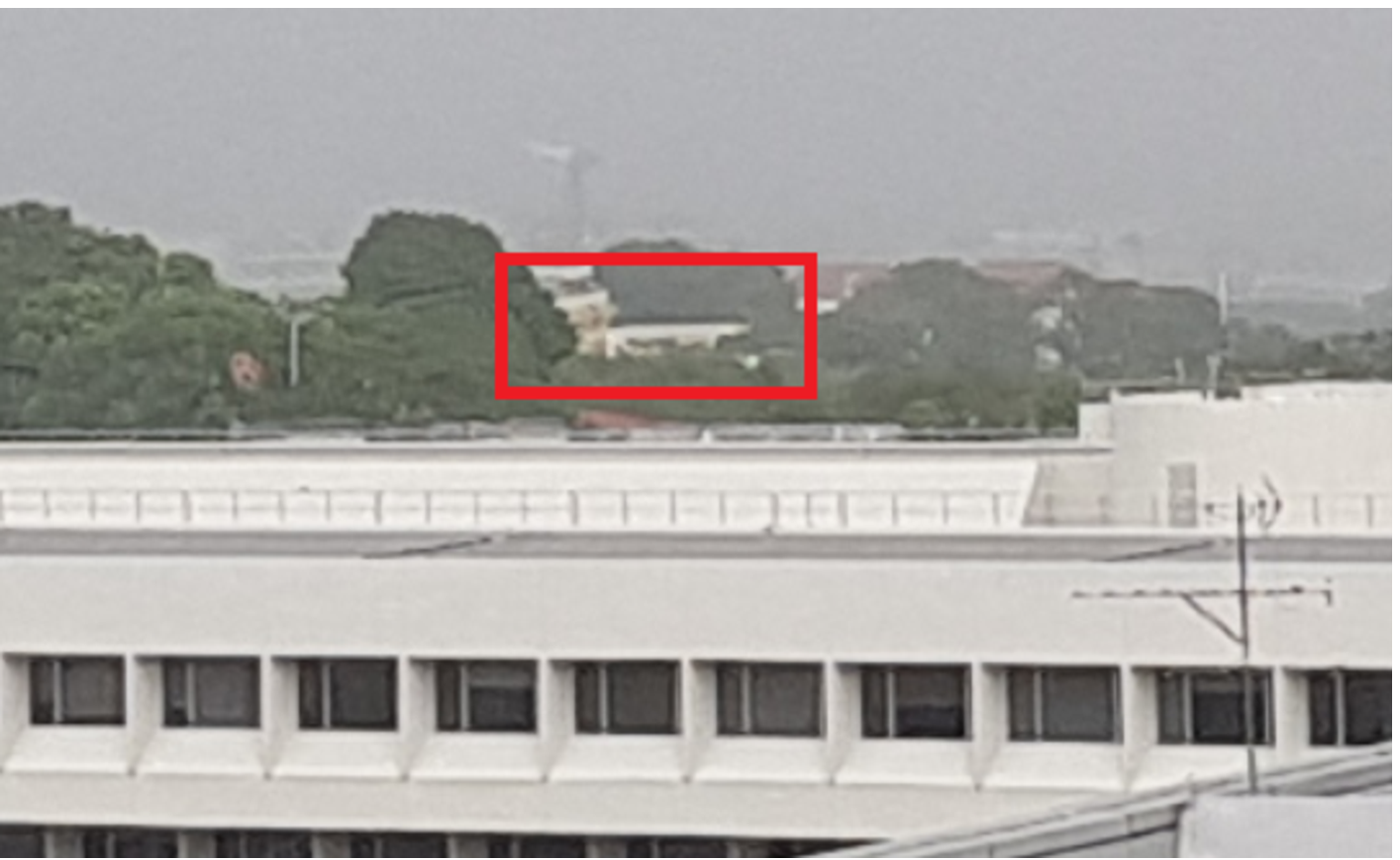}
    \label{fig:a_to_b_small}
  }
  \subfigure[View rooftop from parklot]
  {
    \includegraphics[height=3cm, width=0.29\linewidth]{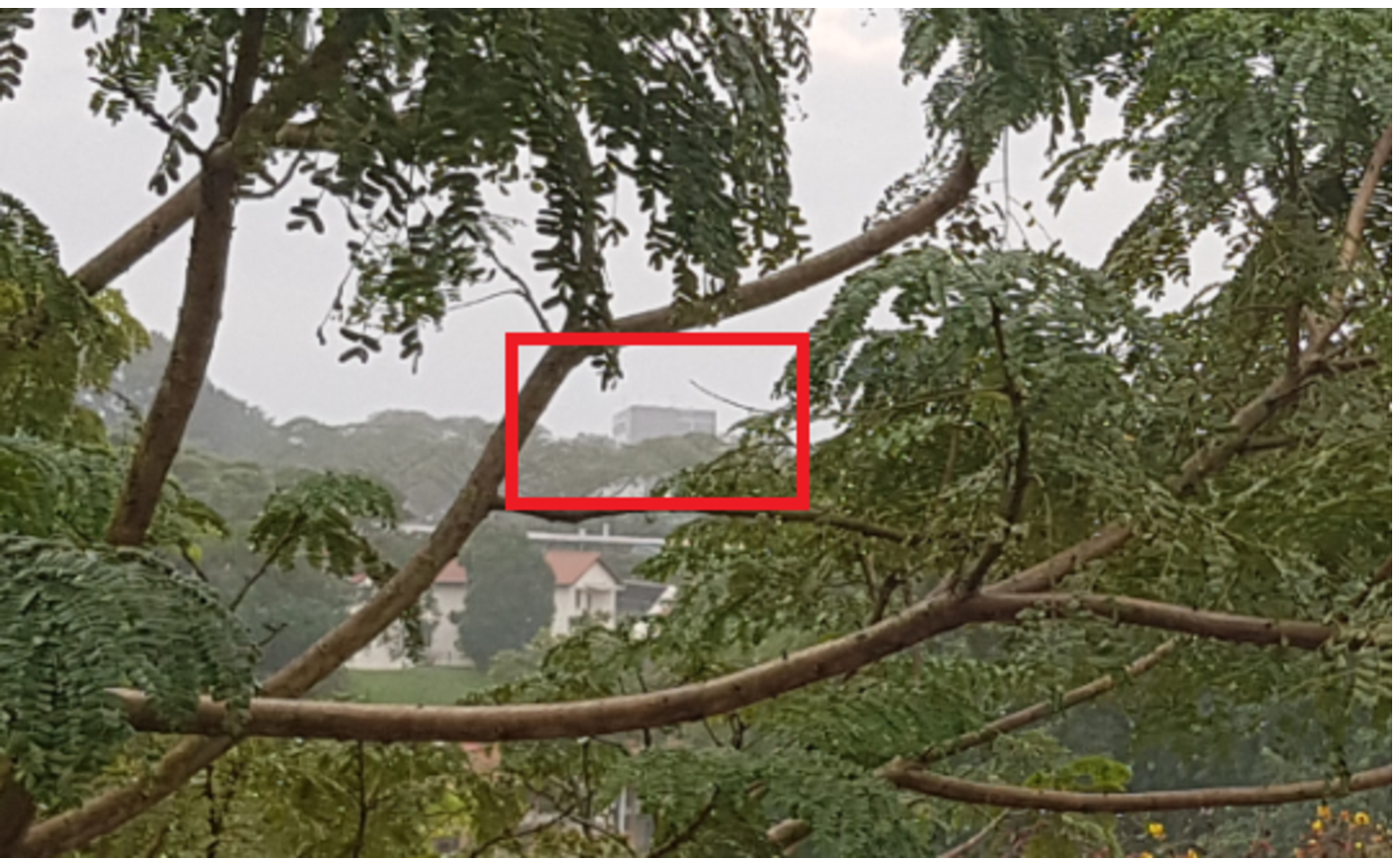}
    \label{fig:b_to_a_small}
  }
  \subfigure[The deployment sites]
  {
    \includegraphics[height=3cm, width=0.29\linewidth]{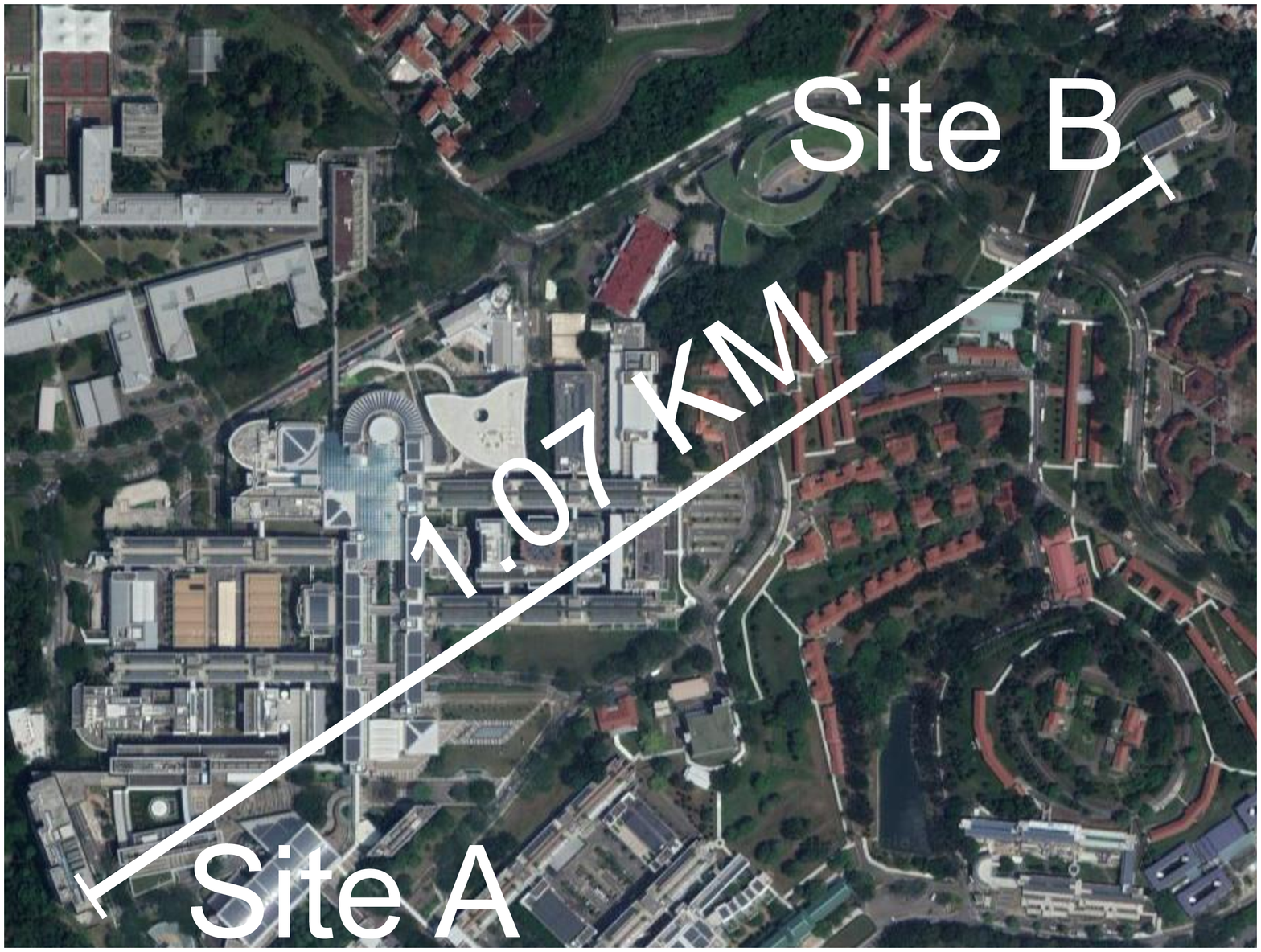}
    \label{fig:campus-map}
  }
  \caption{Pictures taken at the two sites (Site A: rooftop, Site B: parklot).}
  \label{fig:views}
\end{figure}

\begin{figure}
  \centering
  \includegraphics[width=0.9\linewidth]{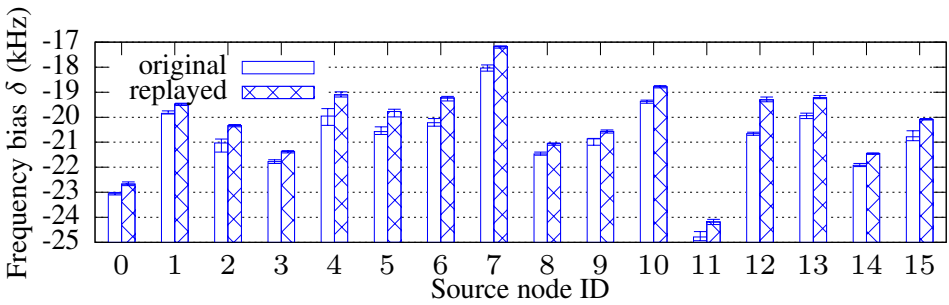}
  \caption{$\mathsf{LoRaTS}$'s FB estimates when the distances between the gateway and the end devices are about $1.07\,\text{km}$.}
  \label{fig:long_dist_fb}
\end{figure}

\noindent {\bf Outdoor experiments with longer distance:}
In this set of experiments, we deploy SX1276-based end devices in an outdoor parking lot. The LoRa parameters for the end devices are the same as we used in the indoor experiments. We replace the iC880a-based gateway shown in Fig.~\ref{fig:real_experiment_detailed} with a $\mathsf{LoRaTS}$ gateway. The distance between the end device and the $\mathsf{LoRaTS}$ gateway is about $1.07\,\text{km}$. 
Fig.~\ref{fig:views} shows the pictures taken at the two sites. The circled construct in a figure is the building where the other site is located in.
The collider shown in Fig.~\ref{fig:real_experiment_detailed} is also used in this set of experiments. The eavesdropper is deployed at a location about $200\,\text{m}$ from the end device. When the transmitting powers of the end device and the collider are $14\,\text{dBm}$ and $8\,\text{dBm}$, respectively, we can successfully launch the frame delay attack. 
We also use the round-trip timing approach discussed in \sect\ref{subsec:accuracy-evaluation} to evaluate AIC's performance. The measurements show that AIC's $\mathrm{RMSD}(\epsilon)$ is $1.29\,\mu\text{s}$ only. This result is better than that obtained in the multistory building because the LoRa signal suffers significant attenuation in the indoor environment. Then, we investigate the additional FBs introduced by the replay attack. Fig.~\ref{fig:long_dist_fb} shows $\mathsf{LoRaTS}$'s FB estimates for the frames transmitted by 16 end devices and the corresponding replays.
The extra FBs introduced by the attack is up to 1.76 ppm. Thus, the attack can be detected.

\vspace{0.2em}
\noindent {\bf Impact of gateway heights on FB estimation:}
\begin{figure}
  \centering
  \includegraphics[width=0.9\linewidth]{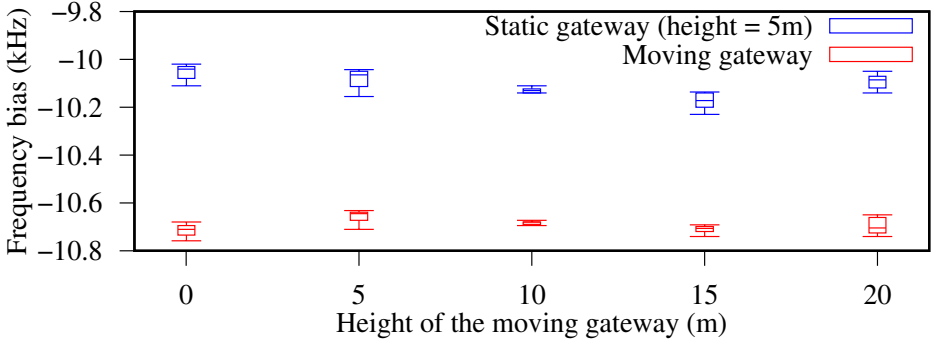}
  \caption{FB estimation results from the \textit{static gateway} and the \textit{moving gateway}. The \textit{static gateway} is placed at a fixed heights ($5m$) while the \textit{moving gateway} is carried to different heights. These two $\mathsf{LoRaTS}$ gateways receive the LoRa signal simultaneously. Each box plot shows minimum, maximum, median, 25\%, and 75\% percentiles.}
  \label{fig:diff_height_fb}
\end{figure}
The device height may affect the communication performance due to the different path losses. To evaluate the impact of gateway height on FB estimation, we use two $\mathsf{LoRaTS}$ gateways to receive the LoRa signal simultaneously in this experiment. We set a {\em static gateway} at a fixed height ($5m$), relative to the end device, and carry the other gateway called {\em moving gateway} to different heights. As shown in Fig.~\ref{fig:diff_height_fb}, the blue box plots and the red box plots show the FB estimation results from the static and moving gateway, respectively. Each box plot shows minimum, maximum, median, 25\%, and 75\% percentiles. From the experiment results, we can see that the gateway's height does not have a noticeable impact on the performance. The FB estimations from the two gateways are different, because the estimated FB contains both the transmitter's and receiver's FBs, as analyzed in \sect\ref{subsubsec:extraction}.

\begin{figure}
  \centering
  \includegraphics[width=0.9\linewidth]{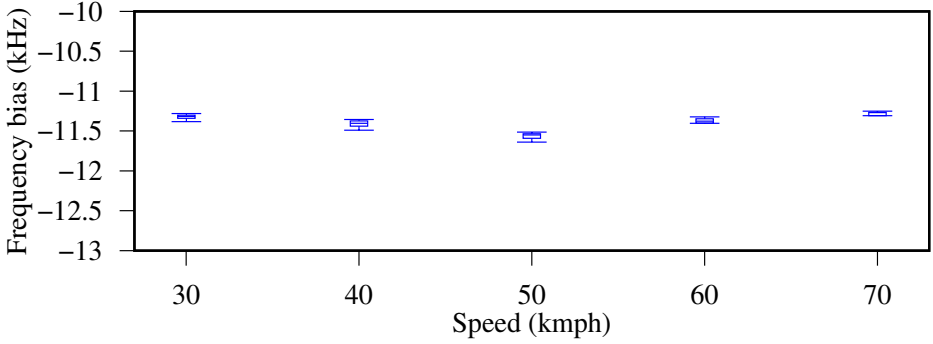}
  \caption{FB estimation results of an end device moving at different speeds. Each box plot shows minimum, maximum, median, 25\%, and 75\% percentiles.}
  \label{fig:diff_speed_fb}
\end{figure}
\vspace{0.2em}
\noindent {\bf Impact of end device movement on FB estimation:}
When there is a relative movement between a transmitter and a receiver, the received signals have frequency shifts caused by the Doppler effect. Previous studies have shown that LoRa modulation is robust against the Doppler effect~\cite{doroshkin2019experimental}. We now analyze the impact of the Doppler effect on the FB estimation. When the end device moves at a velocity $v$ relative to the gateway, the FB caused by the Doppler effect, denoted by $\Delta f_d$, is $\frac{v}{c} f_{chirp}$, where the $c$ is the speed of light and $f_{chirp}$ is the instantaneous frequency of the chirp~\cite{liando2019known}. When the end device moves at a speed of 70 km/h, the additional FB caused by the Doppler effect is about $50$ to $60\,\text{Hz}$.
We conduct a set of experiments to understand the impact of movement speed on the FB estimation. Specifically, the experiments are conducted on a road section. We set up a $\mathsf{LoRaTS}$ gateway at the road side and carry an end device in a car. The car passes the gateway at different speeds of 30, 40, 50, 60, and 70 km/h.
The FB esitmation results are shown in Fig.~\ref{fig:diff_speed_fb}. Each box plot shows minimum, maximum, median, 25\%, and 75\% percentiles of the FB estimates. From the results, we can observe that the FB estimation results for a certain movement speed are stable, with a maximum fluctuation of $145\,\text{Hz}$. Moreover, the FB fluctuation is about $400\,\text{Hz}$ across all speeds. 
This cross-speed fluctuation is caused by the variation of the ambient temperature of the end device during the experiment process.
From our experiment results regarding the impact of temperature on FB that will be presented shortly in \sect\ref{subsec:temp_fb_monitor}, the $400\,\text{Hz}$ FB variation observed in Fig.~\ref{fig:diff_speed_fb} corresponds to a $0.5^\text{o}\text{C}$ temperature variation around $26^\text{o}\text{C}$ (cf. Fig.~\ref{fig:linearity_temp_fb}). Note that $26^\text{o}\text{C}$ is the cabin temperature we set on the car's air conditioning system. As cars' air conditioning systems often have $\pm 1^\text{o}\text{C}$ control errors~\cite{xie2020self}, we cannot completely eliminate the impact of temperature variations in this set of experiments.
In Fig.~\ref{fig:diff_speed_fb}, we do not see the monotonic relationship between FB and movement speed, because the impact of the movement speed (i.e., up to $60\,\text{Hz}$ additional FB according to our analysis) is much smaller than the impact of the temperature variation in the car.

\subsection{Leveraging Temperature for FB Monitoring}
\label{subsec:temp_fb_monitor}
As discussed in \sect\ref{subsec:temporal_stability}, an end device without TCXO needs to transmit frames periodically to help the gateway track its FB in the presence of temperature variations. However, for the end devices that do not have pending application data for transmission, transmitting dummy frames for the purpose of FB tracking wastes energy. To mitigate this issue, we propose to leverage temperature to lower the transmission frequency. In this approach, the end device needs to be equipped with a temperature sensor and attach the temperature reading in the uplink frame. When the uplink frame is received by the $\mathsf{LoRaTS}$ gateway, the gateway will first estimate the FB of the end device. Then, the $\mathsf{LoRaTS}$ estimates the temperature based on a known temperature-FB model and compares the estimated temperature with the actual temperature reading in the payload. If the difference between the estimated value and the sensor reading exceeds a pre-defined threshold, the gateway declares the detection of the stealthy frame delay attack. 

\begin{figure}
  \centering
  \includegraphics[width=\textwidth]{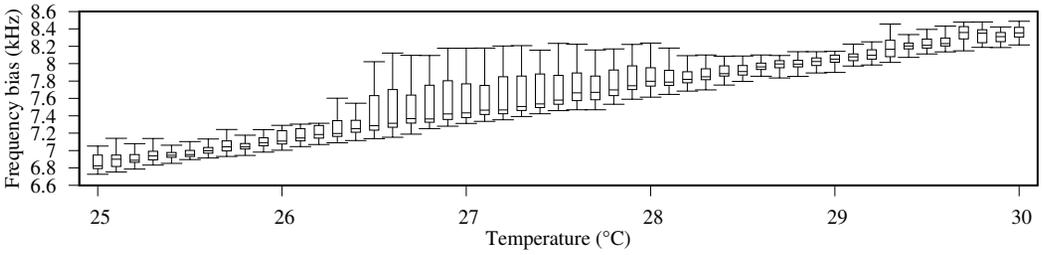}
  \caption{Boxplot of the temperature vs. frequency bias of an SX1276-based end device.}
  \label{fig:linearity_temp_fb}
\end{figure}

\begin{figure}
  \centering
  \includegraphics[width=0.5\textwidth]{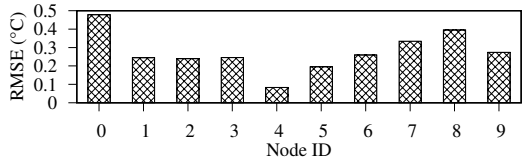}
  \caption{The root-mean-square errors (RMSEs) of linear regression models for 10 SX1276-based end devices.}
  \label{fig:temp-fb-estimation}
\end{figure}
\begin{figure}
  \centering
  \includegraphics[width=\textwidth]{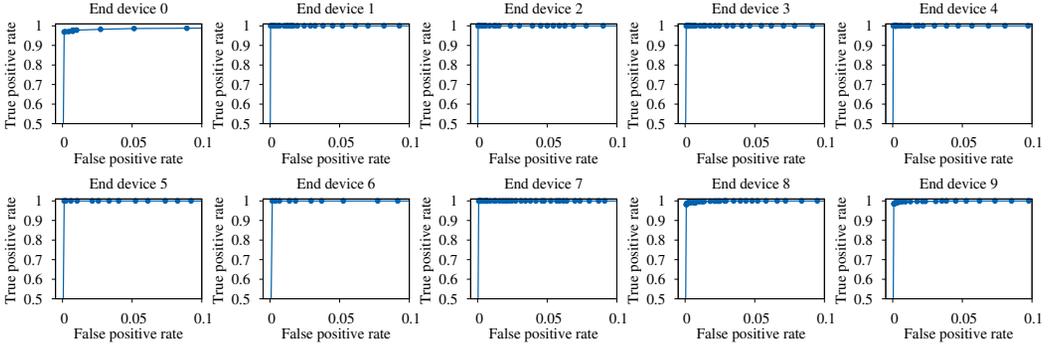}
  \caption{ROCs for linear regression models for 10 end devices.}
  \label{fig:roc}
\end{figure}

To evaluate this approach, we deploy ten SX1276-based end devices, each equipped with an AM2302 temperature-humidity sensor \cite{AM2302}. Each end device transmits a frame, including the reading from the temperature-humidity sensor periodically. Once upon receiving the frame, the gateway timestamps the data, records the temperature, and estimates the FB. Thus, for each end device, we can get a pairwise temperature-FB trace based on collected data. Fig.~\ref{fig:linearity_temp_fb} shows the boxplot of the pairwise data collected by an end device over 24 hours. We can see that the frequency bias increases with the temperature and follows a linear trend. This is because the frequency-temperature characteristics of the crystal oscillator without the temperature compensation exhibit a linear relationship in this temperature range (i.e., $25^\text{o}\text{C}$ to $30^\text{o}\text{C}$) \cite{li2005novel}. Thus, we apply linear regression to model the relationship during the modeling phase and use the model to estimate the temperature according to the FB at run time. We use 2,000 pairwise data points as the training data set in the experiment. Fig. \ref{fig:temp-fb-estimation} shows the root-mean-square errors (RMSEs) of the linear regression models for all the ten end devices. We can see that the RMSEs are below $0.5^\text{o}\text{C}$.
Fig. \ref{fig:roc} shows the receiver operating characteristic (ROC) curves of using the models to detect the frame delay attacks on the frames transmitted by the ten end devices. Different data points on a ROC curve are results based on different temperature detection thresholds. Let $N$ denote the total number of tests for an end device. Accordingly, let $N_{TA}$ and $N_{FA}$ denote the total numbers of true alarms and false alarms at the gateway side, respectively. The true positive rate and the false positive rate are measured by $N_{TA}/N$ and $N_{FA}/N$, respectively. For each end device, our approach achieves empirical true positive rates of 100\%, subject to a false positive rate upper bound of 1\%.

Although this approach needs an additional temperature sensor for each end device, it further lowers the requirement of the communication frequency to save energy in the long term. Note that temperature sensors are often available in many monitoring applications. To capture the runtime affecting factors introduced by different environments and wireless conditions, the network can perform {\em in situ} temperature-FB profiling based on the collected FB and temperature data over a certain time duration. The resulting temperature-FB model thus captures the run-time affecting factors. To prevent the attacker from misleading the gateway to build a false profile, the network operator should perform the profiling in a short attack-free time period with close supervision, e.g., one day in each season. With the closely supervised profiling in a short period of time, we can make sure that the network can mitigate the impact of attack throughout longer periods of time.

\section{Conclusion}
\label{sec:conclude}
This paper shows that sync-free data timestamping, though bandwidth-efficient, is susceptible to the frame delay attack that can be implemented by a combination of frame collision and delayed replay. Experiments show that the attack can affect many end devices in a large geographic area.
To gain attack awareness, we design a gateway called $\mathsf{LoRaTS}$ that integrates a low-power SDR receiver with a commodity LoRaWAN gateway. We develop efficient time-domain signal processing algorithms to estimate the FBs of the end devices. The least squares FB estimation algorithm achieves high resolution and can uncover the additional FBs introduced by the attack. 
We also consider a more skillful and resourceful attacker who eliminates the additional FBs. We propose a pseudorandom interval hopping scheme to counteract the zero-FB attacks.
In summary, with $\mathsf{LoRaTS}$, we can achieve efficient sync-free data timestamping with awareness of frame delay attack.


\begin{acks}
We acknowledge Zhenyu Yan and Dongfang Guo for assistance in conducting the long-distance experiments on the NTU campus. We acknowledge Amalinda J. Gamage and Jansen Christian Liando for building the campus LoRaWAN infrastructure. 
\end{acks}

\bibliographystyle{ACM-Reference-Format}
\bibliography{lora-sync}
\end{document}